\documentclass[preprint,prb,aps,floats,amssymb,showpacs,draft,%
floatfix,superscriptaddress]{revtex4} 
\usepackage{psfig}

\begin{document}

\date{\today}

\title{
The critical exponents of the superfluid transition in $^4$He
}

\author{Massimo Campostrini}
    \affiliation{Dipartimento di Fisica dell'Universit\`a di Pisa
        and I.N.F.N., I-56127 Pisa, Italy}
\author{Martin Hasenbusch}
    \affiliation{Dipartimento di Fisica dell'Universit\`a di Pisa
        and I.N.F.N., I-56127 Pisa, Italy}
\author{Andrea Pelissetto}
    \affiliation{Dipartimento di Fisica dell'Universit\`a di Roma I
        and I.N.F.N., I-00185 Roma, Italy}
\author{Ettore Vicari}
    \affiliation{Dipartimento di Fisica dell'Universit\`a di Pisa
        and I.N.F.N., I-56127 Pisa, Italy}

\begin{abstract}
  We improve the theoretical estimates of the critical exponents for the
  three-dimensional XY universality class, which apply to the superfluid
  transition in $^4$He along the $\lambda$-line of its phase diagram. We
  obtain the estimates $\alpha=-0.0151(3)$, $\nu=0.6717(1)$, $\eta=0.0381(2)$,
  $\gamma=1.3178(2)$, $\beta=0.3486(1)$, and $\delta=4.780(1)$.  Our results
  are obtained by finite-size scaling analyses of high-statistics Monte Carlo
  simulations up to lattice size $L=128$ and resummations of 22nd-order
  high-temperature expansions of two improved models with suppressed leading
  scaling corrections.  We note that our result for the specific-heat exponent
  $\alpha$ disagrees with the most recent experimental estimate
  $\alpha=-0.0127(3)$ at the superfluid transition of $^4$He in microgravity
  environment.
\end{abstract}


\maketitle

\section{Introduction and Summary}
\label{introduction}

The renormalization-group (RG) theory of critical phenomena classifies
continuous phase transitions into universality classes, which are determined
only by a few global properties of the system, such as the space
dimensionality, the nature and the symmetry of the order parameter, the 
symmetry-breaking pattern, and the range
of the interactions.  Within a given universality class, the critical
exponents and scaling functions describing the asymptotic critical behavior
are identical for all systems.  The three-dimensional XY universality class is
characterized by a complex order parameter and symmetry breaking
${\rm O(2)} \cong {\mathbb Z}_2 \otimes {\rm U(1)}\rightarrow{\mathbb Z}_2$.
An interesting representative of this
universality class is the superfluid transition of $^4$He along the
$\lambda$-line $T_\lambda(P)$, which provides an exceptional opportunity for a
very accurate experimental test of the RG predictions, because of the weakness
of the singularity in the compressibility of the fluid and of the purity of the
samples.  Exploiting also the possibility of performing experiments in a
microgravity environment, the specific heat of liquid helium was measured
up to a few nK from the $\lambda$-transition.\cite{Lipa-etal-96}  The
resulting estimate of the specific-heat exponent, obtained after some
re-analyses of the experimental
data,\cite{Lipa-etal-96,Lipa-etal-00,Lipa-etal-03} is \cite{Lipa-etal-03}
\begin{equation}
\alpha=-0.0127(3).
\label{expest}
\end{equation}
Other experimental results at the superfluid transition of $^4$He, and
for other physical systems in the XY universality class, are reported
in Ref.~\onlinecite{PV-rev}.

On the theoretical side the XY universality class has been studied by
various approaches, such as field-theoretical (FT) methods and lattice
techniques based on Monte Carlo (MC) simulations or high-temperature
(HT) expansions.  A review of results can be found in
Ref.~\onlinecite{PV-rev}.  Accurate estimates of the critical exponents were
obtained in Ref.~\onlinecite{CHPRV-01} by combining MC simulations and HT
expansions of improved Hamiltonians. This synergy of lattice
techniques provided the estimate
\begin{equation}
\alpha=-0.0146(8),
\label{theorest}
\end{equation}
which is substantially consistent with the experimental result (\ref{expest}),
although slightly smaller. Other results obtained from FT 
calculations,\cite{GZ-98,ZJ-book,KS-book}
MC simulations,\cite{CHPRV-01,HT-99} and
HT expansions\cite{BC-99,CPRV-00} are less precise and 
in agreement with both 
estimates (\ref{expest}) and (\ref{theorest}).

In this paper we significantly improve the theoretical estimates. This allows
us to check whether the small difference between the theoretical and
experimental estimates (\ref{expest}) and (\ref{theorest}) disappears after a
more accurate theoretical analysis, as recently suggested in
Ref.~\onlinecite{BNPS-05}.  We again follow the strategy of
Ref.~\onlinecite{CHPRV-01}, considering two classes of lattice Hamiltonians,
the $\phi^4$ lattice model and the dynamically diluted XY (ddXY) model. They
depend on an irrelevant parameter, $\lambda$ and $D$ respectively, which can
be tuned to suppress the leading scaling corrections, giving rise to improved
Hamiltonians.  We improve the finite-size scaling (FSS) analysis of these
models by significantly increasing the statistics (by approximately a factor
10) and simulating larger lattices (up to lattice sizes $L=128$).  The
precision of the data allows us to observe the expected next-to-leading
scaling corrections, and therefore to have a much better control of the
systematic errors.  Moreover, we extend the HT expansion of the susceptibility
and of the correlation length in the $\phi^4$ and ddXY models to 22nd order,
i.e., we add two terms to the HT series computed and analyzed in
Refs.~\onlinecite{CHPRV-01,CPRV-00}.  Using this bulk of new data and
calculations, we performed several analyses, also combining information
obtained from MC simulations and IHT analyses (IHT denotes the HT expansion
specialized to improved models).

\begin{figure}[tp]
\centerline{\psfig{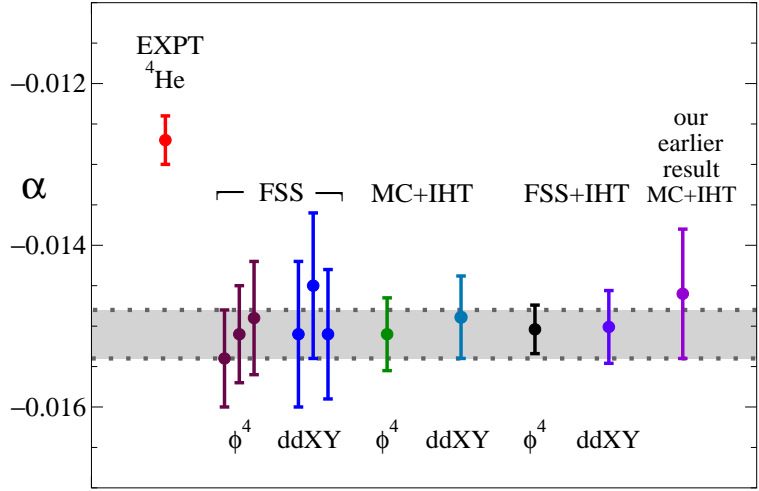}}
\vskip 2mm
\caption{\label{summary}
Summary of our results for the specific-heat exponent $\alpha$. 
Shorthands are explained
in the text. The coloured region corresponds to our final estimate
$\alpha=-0.0151(3)$.
}
\end{figure}

In Fig.~\ref{summary} we show a summary of our results for the specific-heat
critical exponent $\alpha$, as obtained from the hyperscaling relation
$\alpha=2-3\nu$.  We show three sets of new estimates for the $\phi^4$ and
ddXY models, obtained by different methods.  We first report the results of
FSS analyses of MC data up to $L=128$ (FSS).  The three reported results for
each model are obtained from the analysis (left) of a combination of two
quantities (temperature derivatives of RG invariant quantities) that does not
have leading scaling corrections, (center) of the energy density, and (right)
of the fourth-order (Binder) cumulant of the magnetization. Next, we report
the results obtained from the analyses of the 22nd-order IHT series at the
optimal values of the irrelevant parameters $\lambda^*$ and $D^*$, biased
using the MC estimates of $\beta_c$ (MC+IHT). Finally, we show results
obtained by requiring the FSS and IHT analyses to give consistent results for
the exponent $\nu$ (FSS+IHT).  These results come from different analyses of
the available MC data and HT calculations.  Although they are not completely
independent, their comparison represents a highly nontrivial cross-check of
the results and of their errors.  In Fig.~\ref{summary} we also
show our earlier MC+IHT result obtained in Ref.~\onlinecite{CHPRV-01}.

The results shown in Fig.~\ref{summary}
provide a rather accurate estimate of $\alpha$, which we summarize by taking
\begin{equation}
\alpha=-0.0151(3)
\label{newtheorest}
\end{equation}
as our final estimate.  This result significantly improves our earlier
estimate (\ref{theorest}), but it does not reduce the difference from
the experimental result (\ref{expest}). Moreover, now the errors are so
small that their difference appears significant. According to our
analyses, values of the specific-heat exponent $\alpha>-0.014$ appear
to be highly unlikely.  We think that this discrepancy calls for further
investigations.  We mention that a proposal of a new space experiment
has been presented in Ref.~\onlinecite{LWNA-05}.

We also anticipate our best estimates of the other critical exponents:
\begin{eqnarray}
&&\nu=0.6717(1),\label{nu}\\
&&\eta=0.0381(2),\label{eta}\\
&&\gamma=1.3178(2),\label{gamma}\\
&&\delta=4.780(1),\label{delta}\\
&&\beta=0.3486(1).\label{beta}
\end{eqnarray}
Moreover, we obtained an accurate estimate of the exponent $\omega$
associated with the leading scaling corrections, i.e.,
$\omega=0.785(20)$, and of the exponent $\omega_2$ associated with the
next-to-leading scaling corrections, i.e., $\omega_2=1.8(2)$.

The paper is organized as follows.  In Sec.~\ref{models} we define the
$\phi^4$ and ddXY lattice models. Sec.~\ref{FSS} is dedicated to a
summary of the basic RG ideas concerning FSS in critical phenomena. In
Sec.~\ref{FSSanalysis} we present our MC simulations and the FSS
analyses of the data. The computation and analysis of the HT expansion
are presented in Sec.~\ref{HT}.  We report an analysis of the IHT
expansions of improved $\phi^4$ and ddXY models biased using the MC
estimates of $\beta_c$ (MC+IHT), and a combined analysis requiring the
consistency of the IHT and FSS results (FSS+IHT).


\section{Lattice models}
\label{models}

As in Ref.~\onlinecite{CHPRV-01}, we consider two classes of models defined
on a simple cubic lattice and depending on an irrelevant
parameter. They are the $\phi^4$ lattice model and the dynamically
diluted XY (ddXY) model.  The Hamiltonian of the $\phi^4$ lattice
model is given by
\begin{equation}
{\cal H}_{\phi^4} =
 - \beta\sum_{\left<xy\right>} {\vec\phi}_x\cdot{\vec\phi}_y +
   \sum_x \left[ {\vec\phi}_x^{\,2} +
   \lambda ({\vec\phi}_x^{\,2} - 1)^2\right],
\label{phi4Hamiltonian}
\end{equation}
where $\vec{\phi}_x = (\phi_x^{(1)},\phi_x^{(2)})$ is a two-component
real variable. The ddXY model is defined by the Hamiltonian
\begin{equation}
{\cal H}_{\rm dd} =  -  \beta \sum_{\left<xy\right>}   \vec{\phi}_x \cdot
\vec{\phi}_y - D  \sum_x \vec{\phi}_x^{\,2}
\label{ddxy}
\end{equation}
with the local measure
\begin{equation}
d\mu(\phi_x) =  d \phi_x^{(1)} \, d \phi_x^{(2)} \,
\left[
\delta(\phi_x^{(1)}) \, \delta(\phi_x^{(2)})
 + \frac{1}{2 \pi} \, \delta(1-|\vec{\phi}_x|)
\right],
\label{lmeasure}
\end{equation}
and the partition function
\begin{equation}
\int \prod_x d\mu(\phi_x)\, e^{-{\cal H}_{\rm dd}}.
\end{equation}

The parameters $\lambda$ in ${\cal H}_{\phi^4}$ and $D$ in ${\cal
H}_{\rm dd}$ can be tuned to obtain improved Hamiltonians
characterized by the fact that the leading correction to scaling is
absent in the Wegner expansion of any observable near the critical
point.  Considering, for instance, the magnetic susceptibility $\chi$,
the corresponding Wegner expansion is generally given by
\begin{eqnarray}
\chi &=& C t^{-\gamma} \left( 1 + a_{0,1} t + a_{0,2}t^2 + \ldots + a_{1,1}
t^\Delta + a_{1,2} t^{2\Delta} + \ldots \right.
\nonumber \\ 
&& \left. \qquad + b_{1,1} t^{1+\Delta} +
b_{1,2} t^{1+2\Delta} +
\ldots + a_{2,1} t^{\Delta_2} + \ldots \right),
\label{chiwexp}
\end{eqnarray}
where $t\equiv 1 - \beta/\beta_c$ is the reduced temperature. 
We have neglected additional
terms due to other irrelevant operators and 
terms due to the analytic background present in the free 
energy.\cite{AF-83,SS-00,CHPV-02} The leading
exponent $\gamma$ and the correction-to-scaling exponents
$\Delta,\Delta_2,...$, are universal, while the amplitudes $C$,
$a_{i,j}$, $b_{i,j}$ are nonuniversal.  For three-dimensional XY systems, the
value of the leading correction-to-scaling exponent is 
$\Delta\approx0.53$, \cite{CHPRV-01,HT-99,GZ-98} and the value 
of the subleading nonanalytic exponent is $\Delta_2\approx1.2$.\cite{NR-84}
  In the case of improved Hamiltonians the leading correction to
scaling vanishes, i.e., $a_{1,1} = 0$ in Eq.\ (\ref{chiwexp}) (actually
$a_{1,i} = 0$ for all $i$), in the Wegner expansion of any thermodynamic
quantity.

Improved Hamiltonians belonging to the XY universality class have been
discussed in Refs.~\onlinecite{HT-99,CPRV-00,CHPRV-01}.  We mention
that improved Hamiltonians
have also been considered for other universality classes, such as
the Ising \cite{BFMM-98,HPV-99,Hasenbusch-99,CPRV-02} and Heisenberg
universality classes.\cite{Hasenbusch-01,CHPRV-02}

\section{Finite-size scaling}
\label{FSS}

In this section we summarize some basic results concerning FSS in
critical phenomena.  The starting point of FSS is the scaling behavior
of the singular part of the free energy density of a sample of linear
size $L$ (see, e.g., Refs.~\onlinecite{SS-00,Privman-90}):
\begin{eqnarray}
{\cal F}_{\rm sing} (u_t, u_h, \{u_i\},L)  =
 L^{-d}
{\cal F}_{\rm sing}( L^{y_t} u_t, L^{y_h} u_h, \{L^{y_i} u_i\}),
\label{FscalL}
\end{eqnarray}
where $u_t\equiv u_1$, $u_h\equiv u_2$, $\{u_i\}$ with $i\geq 3$ are
the scaling fields (which are analytic functions of the Hamiltonian
parameters) associated respectively with the reduced temperature $t$
($u_t\sim t$), magnetic field $H$ ($u_h\sim H$), and the other
irrelevant perturbations with $y_i<0$.  The scaling behavior of the
interesting thermodynamic quantities can be obtained by performing
the appropriate derivatives of Eq.~(\ref{FscalL}), with respect to $t$
and $H$.  Since $u_t$ and $u_h$ are assumed to be the only relevant
scaling fields, one may expand with respect to the arguments
corresponding to the irrelevant scaling fields. This provides the
leading scaling behavior and the power-law scaling corrections.

The RG exponents of the relevant scaling fields $u_t$ and $u_h$ are
related to the standard exponents $\nu$ and $\eta$, i.e., $y_t=1/\nu$
and $y_h=(d+2-\eta)/2$.  The RG exponent $y_3\equiv-\omega$ of the
leading irrelevant scaling field $u_3$ has been estimated by the
analysis of high-order FT perturbative expansions,\cite{GZ-98}
obtaining $\omega=0.802(18)$ ($\epsilon$ expansion) and
$\omega=0.789(11)$ ($d=3$ expansion).  Results from lattice techniques
are in substantial agreement, see Ref.~\onlinecite{PV-rev}. As
we shall see later, our FSS analysis provides the estimate
$\omega=0.785(20)$.  Concerning the next-to-leading scaling
corrections, we mention the FT results of Ref.~\onlinecite{NR-84}: $y_4
=-1.77(7)$ and $y_5=-1.79(7)$ ($y_{421}$ and $y_{422}$ in their
notation).  Note that, at present, there is no independent check of
these results. As we shall see, our new MC and HT analyses confirm
that the next-to-leading scaling corrections are characterized by a RG
exponent $y_4=-1.8(2)$.  There are also corrections due to the
violation of rotational invariance by the lattice; the corresponding
RG dimension is $y_6=-2.02(1)$.\cite{CPRV-98}  We mention that for
some quantities, such as the Binder cumulant and the ratio
$R_\xi\equiv \xi/L$, there are also corrections due to the analytic
background of the free energy.  This should lead to
corrections with $y_7=-(2-\eta) = -1.9619(2)$ (obtained by using the
estimate of $\eta$ of this work). Finally, 
in the case of $R_\xi$, we expect also $O(L^{-2})$ corrections,\cite{CP-98}
related to the particular definition (\ref{xidef}) of
$\xi$.  

For vanishing external field $H$,
the behavior of a phenomenological
coupling $R$, i.e., of a quantity that is invariant under RG
transformations in the critical limit, can be written in the FSS limit as
\begin{equation}
R(L,\beta,\lambda) =
r_0(u_t L^{y_t}) + \sum_k r_{3,k}(u_t L^{y_t}) \, u_3^k L^{k y_3} + 
\sum_{i\ge4} r_i(u_t L^{y_t}) \, u_i L^{y_i} + \ldots \, ,
\label{Rexp_1}
\end{equation}
where we have singled out the corrections due to the 
leading irrelevant operator.
The functions
$r_0(z)$, $r_i(z)$, $r_{3,k}(z)$
are smooth and finite for $z\to0$ and, by definition,
$u_t(\beta,\lambda)\sim t\equiv 1 - \beta/\beta_c$. 
In general, all others scaling fields 
$u_i$ are finite for $t = 0$. Improved models are characterized by the 
additional condition that $u_3(t = 0) = 0$: in this case, all 
corrections proportional to $L^{k y_3} = L^{-k \omega}$ vanish at the 
critical point $t=0$.
In the limit $t\to 0$
and $u_t L^{y_t}\sim t L^{1/\nu}\to 0$, we can further
expand Eq.\ (\ref{Rexp_1}), obtaining
\begin{equation}
\label{expandR}
R(L,\beta,\lambda) =
R^* + c_t(\beta,\lambda) \, t L^{y_t}
  + \sum_i  c_i(\beta,\lambda) \, L^{y_i}
  + O[t^2 L^{2 y_t}, L^{2 y_3}, t L^{y_t+y_3}],
\end{equation}
where $R^*\equiv r_0(0)$. As we already mentioned, in improved models
all corrections proportional to $L^{k y_3}$ vanish for $t = 0$.

Instead of computing the various quantities at fixed Hamiltonian
parameters, one may study the FSS keeping a phenomenological
coupling $R$ fixed at a given value $R_{f}$.  This means
that, for each $L$, one considers 
$\beta_f(L)$ such that
\begin{equation}
R(L,\beta=\beta_f(L)) = R_{f}.
\label{rcbeta}
\end{equation}
All interesting thermodynamic quantities are then measured at $\beta =
\beta_f(L)$.  The pseudocritical temperature $\beta_f(L)$ converges to
$\beta_c$ as $L\to \infty$.  The value $R_{f}$ can be specified at will, as
long as $R_f$ is taken between the high- and low-temperature fixed-point
values of $R$.  The choice $R_{f} = R^*$ (where $R^*$ is the
critical-point value) improves the convergence of $\beta_f$ to $\beta_c$ for
$L\to\infty$;  indeed~\cite{Hasenbusch-99,CHPRV-01}
$\beta_f-\beta_c=O(L^{-1/\nu})$ for generic values of $R_f$, while
$\beta_f-\beta_c=O(L^{-1/\nu-\omega})$ for $R_f=R^*$.  This method has several
advantages. First, no precise knowledge of $\beta_c$ is needed. Secondly, for
some observables, the statistical error at fixed $R_f$ is smaller 
than that  at fixed $\beta = \beta_c$.

Typically, the thermal RG exponent $y_t=1/\nu$ is computed from the
FSS of the derivative of a phenomenological coupling $R$ with respect
to $\beta$ at $\beta_c$.  Using Eq.\ (\ref{Rexp_1}) one obtains
\begin{equation}
\label{numethod}
\left. S_R \right |_{\beta_c} \equiv
\left. \frac{\partial R}{\partial\beta} \right|_{\beta_c} =
-\frac{1}{\beta_c}\left[ 
r_0'(0)\,L^{y_t} + \sum_{i=3} r'_i(0)\,u_i(\beta_c)\,L^{y_i + y_t} +
   \sum_{i=3} r_i(0)\,u'_i(\beta_c)\,L^{y_i} + ... \right] .
\label{scaling-derivative-R}
\end{equation}
An analogous expansion holds for $S_R$ computed at a  fixed
phenomenological coupling.
The leading corrections scale with $L^{y_3}$.  However, in
improved models in which $u_3(\beta_c) = 0$, the leading correction is
of order $L^{y_4}$.  Note that corrections proportional to
$L^{y_3-y_t} \approx L^{-2.3}$ are still present even if the model is
improved.
If one computes the derivative at $\beta_c$, one should also take into account
the uncertainty on $\beta_c$ in the error estimate of $\nu$;
therefore, also in this case it is more
convenient to evaluate $S_R$ at $\beta_f$ defined in 
Eq.\ (\ref{rcbeta}).

Before concluding the section, we would like to recall three basic
assumptions of FSS when considering boundary conditions consistent
with translation invariance, such as periodic boundary
conditions:\cite{SS-00}
\begin{itemize}
\item[(a)]
$1/L$ is an exact scaling field with no corrections proportional
to $1/L^2$, $1/L^3$, etc.;
\item[(b)]
the nonlinear scaling fields have coefficients that are $L$ independent;
\item[(c)]
the analytic background present in the free energy 
depends on $L$ through exponentially small terms.
\end{itemize}

The theoretical evidence for these three hypotheses is discussed in
Ref.~\onlinecite{SS-00}.  Under these assumptions, there are no
analytic $1/L$ corrections.  These assumptions can be verified
analytically in the two-dimensional Ising model, see, e.g.,
Refs.~\onlinecite{CHPV-02,Salas-02,IH-02} and references therein.  As far as
we know, all numerical results reported in the literature are in full
agreement with assumptions (a), (b), and (c).  In particular, there is
no evidence of $1/L$ corrections to FSS.  As we shall
see, the FSS analysis that we shall present in this paper will provide
further support to the FSS assumptions, and in particular to the
absence of $1/L$ analytic corrections.

\section{Monte Carlo simulations and finite-size scaling analyses}
\label{FSSanalysis}

In this section we present MC simulations that significantly
extend those of Ref.~\onlinecite{CHPRV-01}. The
statistics are much larger and we consider larger lattice sizes 
(the largest lattice has $L = 128$).  As we
shall see, the new MC data allow us to perform a more accurate FSS
analysis, achieving a much better control of the next-to-leading
scaling corrections, and therefore of the systematic errors
related to the subleading scaling corrections.

\subsection{Monte Carlo simulations}
\label{MCsim}

We simulated the $\phi^4$ and ddXY models on simple cubic lattices of
size $L^3$ with periodic boundary conditions, at several values of the
Hamiltonian parameters $\lambda$ and $D$, close to 
the optimal values $\lambda^*$ and $D^*$, at which 
leading scaling corrections vanish. Most of the simulations
correspond to  $\lambda=2.07$ and
$D= 1.02$, that represent the best estimates of 
$\lambda^*$ and $D^*$ of Ref.~\onlinecite{CHPRV-01}.

The basic algorithm is the same as in our previous numerical study reported in
Ref.~\onlinecite{CHPRV-01}. We use a combination of local and
cluster\cite{Wolff-89} updates.  We perform wall-cluster updates:\cite{HPV-99}
we flip all clusters that intersect a plane of the lattice.  A cluster update
changes only the angle of the variables. An ergodic algorithm is achieved by
adding local updates,\cite{BT-89} which can also change the length of the spin
variables. We use Metropolis and overrelaxation algorithms
as local updates, which we alternate with the cluster updates.

The main difference with respect to our previous numerical work
\cite{CHPRV-01} concerns the random number generator.  Since we planned to
increase the statistics by approximately one order of magnititude, we decided
to use a higher-quality random number generator, such as those proposed in
Ref.~\onlinecite{ranlxd}.  In our MC simulations we used the {\tt ranlux}
random number generator with luxury level 2.  Its main drawback is that it
requires much more CPU time than the G05CAF generator of the 
NAG-library~\cite{G05CAFgen} which we  
used in Ref.~\onlinecite{CHPRV-01}.  In order to get a good performance, despite the
use of the expensive random number generator, we used demonized versions
\cite{Creutz-83,Creutz-92} of the update algorithms, which allowed us to save
many random numbers.

Most simulations of the $\phi^4$ and ddXY models were performed at the
estimates of $\lambda^*$ and $D^*$ obtained in Ref.~\onlinecite{CHPRV-01},
i.e.,  $\lambda=2.07$ and $D=1.02$. In addition, we perfomed simulations
at $\lambda=1.9,2.1,2.2,2.3$ for the
$\phi^4$ model, and at $D=0.9,1.2$ for the ddXY model.  We shall
also present some MC simulations of the standard XY model.
In total, the MC simulations took approximately 20 years of CPU-time on a
single 2.0 GHz Opteron processor.

\subsection{Definitions of the measured quantities}
\label{def}

The energy density is defined as
\begin{equation}
\label{energy}
E=\frac{1}{V} \sum_{\left<xy\right>} \vec{\phi}_x \cdot \vec{\phi}_y\; .
\end{equation}
The magnetic susceptibility $\chi$ and the correlation length $\xi$
are defined as
\begin{equation}
\chi  \equiv  \frac{1}{V} \,
\biggl\langle \Big(\sum_x \vec{\phi}_x \Big)^2 \biggr\rangle
\end{equation}
and
\begin{equation}
\xi  \equiv  \sqrt{\frac{\chi/F-1}{4 \sin^2 \pi/L}},
\label{xidef}
\end{equation}
where
\begin{equation}
F  \equiv  \frac{1}{V} \, \biggl\langle
\Big|\sum_x \exp\left(i \frac{2 \pi x_1}{L} \right)
        \vec{\phi}_x \Big|^2
\biggr\rangle
\end{equation}
is the Fourier transform of the correlation function at the lowest
non-zero momentum.

We also consider several so-called phenomenological couplings, i.e.,
quantities that, in the critical limit, are invariant under RG
transformations. We
consider the Binder parameter $U_4$ and its sixth-order generalization
$U_6$, defined as
\begin{equation}
U_{2j} \equiv \frac{\langle(\vec{m}^2)^j\rangle}{\langle\vec{m}^2\rangle^j},
\end{equation}
where $\vec{m} = \frac{1}{V} \, \sum_x \vec{\phi}_x$ is the magnetization of
the system.  We also consider the ratio $R_Z\equiv Z_a/Z_p$ of the partition
function $Z_a$ of a system with antiperiodic boundary conditions in one of the
three directions and the partition function $Z_p$ of a system with periodic
boundary conditions in all directions.  Antiperiodic boundary conditions in
the first direction are obtained by changing the sign of the term
$\vec{\phi}_x \cdot \vec{\phi}_y$ of the Hamiltonian for links
$\left<xy\right>$ that connect the boundaries, i.e., for $x=(L,x_2,x_3)$ and
$y=(1,x_2,x_3)$.  Finally, we define the helicity modulus $\Upsilon$.  For
this purpose we introduce a twisted term in the Hamiltonian.  More precisely,
we consider the nearest-neighbor sites $(x, y)$ with $x_1=L$, $y_1=1$,
$x_2=y_2$ and $x_3=y_3$, and replace the term $\vec{\phi}_{x}\cdot
\vec{\phi}_{y}$ in the Hamiltonian with
\begin{equation}
 \vec{\phi}_{x} \cdot R_{\varphi} \vec{\phi}_{y} =
 \phi_{x}^{(1)} \left(\phi_{y}^{(1)} \cos\varphi
                         + \phi_{y}^{(2)} \sin\varphi  \right)+
 \phi_{x}^{(2)} \left(\phi_{y}^{(2)} \cos\varphi
                         - \phi_{x}^{(1)} \sin\varphi \right),
\label{ydef1}
\end{equation}
where $R_{\varphi}$ is a rotation by an angle $\varphi$.  The helicity
modulus is defined by
\begin{equation}
\label{ups_infffxy}
\Upsilon \equiv  -  {1\over L} \left .
\frac{\partial^2 \ln Z(\varphi)}{\partial \varphi^2} \right |_{\varphi=0} .
\end{equation}
The quantities $U_{2j}$, $R_Z\equiv Z_a/Z_p$, $R_\xi\equiv \xi/L$, and
$R_\Upsilon\equiv \Upsilon L$ are invariant under RG transformations
in the critical limit. Thus, they can be considered as
phenomenological couplings. In the following we will generally refer
to them by using the symbol $R$.  Finally, we also consider the derivative
of the phenomenological couplings with respect to the
inverse temperature, i.e.,
\begin{equation}
S_R \equiv { \partial R\over \partial \beta},
\label{sdef}
\end{equation}
which allows us to determine the critical exponent $\nu$
through Eq. (\ref{numethod}).

\subsection{Determination of the critical temperature}

As a first step of the analysis we determine the inverse transition
temperature $\beta_c$ for various values of $\lambda$ in the $\phi^4$
model and of $D$ in the ddXY model. For this purpose we employ the
standard Binder-crossing method, i.e.,  $\beta_c$ is determined by requiring
\begin{equation}
\label{Rsfit}
R(\beta_c,L) = R^*
\end{equation}
independently of $L$. 
Here $R$ is a phenomenological coupling,
$R^*$ is its fixed-point value, and corrections to scaling
are ignored.  In practice we compute $R(\beta,L)$ as Taylor series up
to third order around the simulation value $\beta_{\rm run}$.
We choose as $\beta_{\rm run}$ our previous best estimate of $\beta_c$, see 
Ref.~\onlinecite{CHPRV-01}.  In the analysis we consider four different
phenomenological couplings:
$R_Z\equiv Z_a/Z_p$, $R_\xi\equiv \xi/L$, $U_4$, and $U_6$.  We
do not use the helicity modulus because it has a poor statistical
accuracy for large lattices, which are important to determine
$\beta_c$.

\begin{table}
\squeezetable
\caption{
Estimates of $\beta_c$ at $\lambda=2.07$ and $D=1.02$ and of the 
fixed-point value $R^*$ 
for several dimensionless quantities. Results of 
fits of the MC data in the range $48 \le L \le 128$. 
For details, see the text.  }
\label{tabRstar}
\begin{ruledtabular}
\begin{tabular}{cccc}
\multicolumn{1}{c}{quantity}&
\multicolumn{1}{c}{$\beta_c$ at $\lambda=2.07$}&
\multicolumn{1}{c}{$\beta_c$ at $D=1.02$}&
\multicolumn{1}{c}{$R^*$} \\
\colrule
  $R_Z$   &  0.5093835(2)[3]  &   0.5637963(2)[2]  &  0.3203(1)[3] \\
 $R_\xi$  &  0.5093836(2)[3]  &   0.5637963(2)[2]  &  0.5924(1)[3] \\
    $U_4$     &  0.5093833(3)[1]  &   0.5637961(3)[1]  &  1.2431(1)[1] \\
    $U_6$     &  0.5093834(3)[2]  &   0.5637962(3)[2]  &  1.7509(2)[7] \\
\end{tabular}
\end{ruledtabular}
\end{table}

We first analyze the data at $\lambda=2.07$ for the $\phi^4$ model and
$D=1.02$ for the ddXY model, for which we have data with higher statistics
especially for the largest lattices.  We perform fits with ansatz
(\ref{Rsfit}) for the two models separately.  The two models provide
consistent results for $R^*$, as required by universality.  In order to
improve the statistical accuracy, we also perform joint fits of the results 
for both models,
imposing the same value of $R^*$.  We perform fits with various $L_{\rm min}$
and $L_{\rm max}$ (respectively the smallest and largest lattice size taken
into account), to check stability.  In Table \ref{tabRstar} we report our
final results, which are taken from joint fits of the results for the two models
($\lambda=2.07$ and $D=1.02$) with $48 \le L \le 128$.  Systematic errors are
estimated by comparison with fits with $24\le L \le 48$, i.e., by
evaluating (difference of the two fits)$/(2^x-1)$ with
$x=1/\nu+\omega \approx 2.3$ for $\beta_c$ and $x=\omega \approx 0.8$ for
$R^*$ (here we pessimistically assume that leading corrections dominate).  The
estimates of $\beta_c$ obtained by using different quantities are all 
consistent among each other. As our final result we take the one obtained 
from the data
of $R_Z$: $\beta_c = 0.5093835(2)[3]$ for the $\phi^4$ model at $\lambda=2.07$
and $\beta_c=0.5637963(2)[2]$ for the ddXY model at $D=1.02$,
where the number in parentheses is the statistical error, while the
number in brackets is the systematic error due to scaling corrections.
In Table~\ref{tabRstar} we also report the estimates of the fixed-point values
$R^*$ of the phenomenological couplings, which improve the results of
Ref.~\onlinecite{CHPRV-01}.
For the other values of $\lambda$ and $D$ considered,
we estimate $\beta_c$ by requiring that $R(\beta_c,L=128)=R^*$;
for $R^*$ we use 
the estimate of $R^*$ reported in Table~\ref{tabRstar}. Again, the
best estimate is obtained from $R_Z$; for this quantity
scaling corrections are quite
small.  The results are reported in Table \ref{betac}, where the
number in parentheses is the statistical error, while the
number in brackets is the error due to the uncertainty on $R^*$.

\begin{table}
\squeezetable
\caption{
Estimates of $\beta_c$ for several values of $\lambda$ and $D$, from the
FSS analysis of the MC data, and from the fit of MC data of $\chi$ and
$\xi$ in the HT phase using bIA1 approximants of the 22nd-order HT
series (MC-HT) of $\chi$ and $\xi$, see Sec.~\ref{resht}. The results for 
$\lambda = 2.00$ ($\phi^4$ model) and $D = 0.90, 1.03$ (ddXY model)
are taken from Ref.~\protect\onlinecite{CHPRV-01}.
}
\label{betac}
\begin{ruledtabular}
\begin{tabular}{llll}
\multicolumn{1}{c}{model}&
\multicolumn{1}{c}{FSS}&
\multicolumn{1}{c}{MC-HT from $\chi$}&
\multicolumn{1}{c}{MC-HT from $\xi$}\\
\colrule
$\phi^4,\,\lambda=1.90$ & 0.5105799(4)[3] & &\\
$\phi^4,\,\lambda=2.00$ & 0.5099049(15)   & &  \\
$\phi^4,\,\lambda=2.07$ & 0.5093835(2)[3] & &\\
$\phi^4,\,\lambda=2.10$ & 0.5091503(3)[3] & 0.5091504(4)& 0.5091504(4) \\
$\phi^4,\,\lambda=2.20$ & 0.5083355(3)[4] & 0.5083361(4)& 0.5083363(4) \\
ddXY,$\,D=0.90$   & 0.5764582(15)[9] & & \\
ddXY,$\,D=1.02$   & 0.5637963(2)[2] & 0.5637956(6) & 0.5637970(7) \\
ddXY,$\,D=1.03$   & 0.5627975(7)[7] & & \\
ddXY,$\,D=1.20$   & 0.5470376(17)[6] & 0.5470383(6) & 0.5470392(7) \\
standard XY & 0.4541652(5)[6]%
\footnote{This estimate is obtained
from a fit with  $R_Z(\beta_c,L) = R_Z^* + c L^{-\omega}$,
with $L_{\min}=32$, fixing $R_Z^*=0.3203$ and $\omega=0.785$.
This result is consistent with $\beta_c =0.4541659(10)$ given in
Ref.~\onlinecite{bloete}.}
 & & \\
\end{tabular}
\end{ruledtabular}
\end{table}

\subsection{FSS at fixed phenomenological coupling $R$}
\label{fssfixed}

Instead of performing the FSS analysis at fixed Hamiltonian parameters, one
may analyze the data at a fixed value of a given phenomenological coupling
$R$, as discussed in Sec.~\ref{FSS}. For this purpose we need to compute
$R(\beta)$ in a neighborhood of $\beta_c$.  This could be done by reweighting
the MC data obtained in a simulation at $\beta = \beta_{\rm run} \approx
\beta_c$. However, due to our enormous statistics, we could not store all
results needed to reweight the data. Instead, we computed the first
derivatives of $R(\beta)$ with respect to $\beta$ and determined $R(\beta)$ by
using its third-order Taylor expansion around $\beta_{\rm run}$. We checked
that this is by far enough for our purpose.  If we do not use the reweighting
technique, it is enough to store bin-averages of the different quantities,
significantly reducing the amount of needed disk space.  Given $R(\beta)$, one
determines the value $\beta_{f}$ such that $R(\beta = \beta_f) = R_{f}$.  All
interesting observables are then measured at $\beta_f$; their errors at fixed
$R=R_f$ are determined by a standard jackknife analysis.  For compatibility
with our previous study,\cite{CHPRV-01} we choose $R_{Z,f} =0.3202$ and
$R_{\xi,f}=0.5925$.

This method has the advantage that it does not require a precise
knowledge of the critical value $\beta_c$. But there is another nice
side effect: for some observables the statistical errors at fixed
$R_f$ are smaller than those at fixed $\beta$ (close to $\beta_c$).
This is due to cross-correlations and to a reduction of the effective
autocorrelation times. For example, we find
\begin{eqnarray}
{\mbox{err}[\chi|_{\beta_c}] \over \mbox{err}[\chi|_{R_Z=0.3202}] }
    \approx 3.2,    &\qquad&
{\mbox{err}[\chi|_{\beta_c}] \over \mbox{err}[\chi|_{R_\xi=0.5925}]}
    \approx 4.5,    \\
{\mbox{err}[U_4|_{\beta_c}]\over \mbox{err}[U_4|_{R_Z=0.3202}]}
    \approx 1.9,    &\qquad&
{\mbox{err}[U_4|_{\beta_c}]\over\mbox{err}[U_4|_{R_\xi=0.5925}]}
    \approx 1.6    \nonumber
\end{eqnarray}
for the ddXY model, with a very small $L$ dependence (within the
last figure of the above-reported numbers).  In the case of the $\phi^4$ model
we find slightly smaller improvements for the same quantities.  We
also mention that the gain is marginal for the derivatives of
$R$ considered in this paper.  A reduction
of the statistical errors when some quantities are measured at a fixed
$R$ has also been observed in other models.\cite{HPV-05}

\subsection{The leading correction-to-scaling exponent $\omega$}
\label{omega}

In order to study corrections to scaling we analyze the value of a
phenomenological coupling $R_1$ at a fixed value of a second coupling
$R_2$. If $\beta_f$ is determined by
$R_2(\beta_f)=R_{2,f}$, we consider
\begin{equation}
\bar{R}_1 \equiv R_1(\beta_f)
\label{barr}
\end{equation}
(the dependence on $L$ is understood hereafter).
Note that the large-$L$ limit of
$\bar{R}_1$ is universal but depends on $R_{2,f}$. It differs
from the critical value $R_1^*$, unless $R_{2,f}=R_2^*$.  The
phenomenological couplings that we consider are $U_4$, $U_6$, $R_Z\equiv
Z_a/Z_p$, $R_\xi\equiv \xi/L$, and $R_\Upsilon\equiv \Upsilon L$.

\begin{table}
\squeezetable
\caption
{ Fits to ansatz~(\ref{correctionfit}), using the data at
$\lambda_1=1.9$ and $\lambda_2=2.3$ in the case of the $\phi^4$ model
and $D_1=0.9$ and $D_2=1.2$ in the case of the ddXY model. $L_{\rm
min}$ and $L_{\rm max}$ are the minimal and maximal lattice size that
has been included in the fit. 
$\chi^2/{\rm d.o.f.}$ is the chi square per degree of freedom of 
the fit. }
\label{tabcorr1}
\begin{ruledtabular}
\begin{tabular}{lccrrllc}
\multicolumn{1}{c}{model}&
\multicolumn{1}{c}{$R_1$}&
\multicolumn{1}{c}{$R_2$}&
\multicolumn{1}{c}{$L_{\rm min}$}&
\multicolumn{1}{c}{$L_{\rm max}$}&
\multicolumn{1}{c}{$\Delta c$}&
\multicolumn{1}{c}{$\omega$}&
\multicolumn{1}{c}{$\chi^2/{\rm d.o.f.}$}\\
\colrule
$\phi^4$ & $U_4$ & $R_Z$ & 5 & 12  & $-$0.0209(2) & 0.825(4) &  0.8 \\
& & & 10 & 24 & $-$0.0200(5) & 0.804(10) & 1.1 \\

& $U_4$ & $R_\xi$ & 5 & 12  &$-$0.0210(2) & 0.775(4) & 0.5 \\
& & & 10 & 24  & $-$0.0215(5) & 0.785(10)& 1.4 \\

& $R_\Upsilon$ & $R_Z$ & 5 &12& $-$0.0053(1)&0.722(9) &1.4 \\
& & &10 &24& $-$0.0060(5) & 0.775(37)& 0.7 \\

ddXY & $U_4$ & $R_Z$ & 5 & 12 & $-$0.0355(2) & 0.782(3) & 0.7  \\
& & & 10& 24 & $-$0.0349(6)  & 0.775(7) & 0.9 \\

& $U_4$  & $R_\xi$ & 5 & 12& $-$0.0365(2) & 0.739(3) & 4.3 \\
& & & 10 & 24& $-$0.0386(7)& 0.764(7) & 0.6 \\

& $R_\Upsilon$ & $R_Z$ & 5 & 12 & $-$0.0099(1) &0.708(5) &3.6 \\
& & & 10 & 24& $-$0.0115(7)& 0.773(25) & 0.9 \\
\end{tabular}
\end{ruledtabular}
\end{table}

In the $\phi^4$ model we define
\begin{equation}
\label{deltadef}
 \Delta(\lambda_1,\lambda_2) \equiv \bar{R}(\lambda_2) - \bar{R}(\lambda_1),
\end{equation}
and analogously for the ddXY model, replacing $\lambda$ with $D$.
Since
\begin{equation}
\bar{R}(\lambda) = \bar{R}^* + c(\lambda) L^{-\omega} + \ldots,
\end{equation}
we perform fits with the most simple ansatz
\begin{equation}
\label{correctionfit}
\Delta(\lambda_1,\lambda_2) = \Delta c L^{-\omega},
\qquad \Delta c = c(\lambda_2) - c(\lambda_1),
\end{equation}
and analogously for the ddXY model.  A selection of such fits is given
in Table~\ref{tabcorr1}, where we report only results for $\bar{U}_4$ and
$\bar{R}_\Upsilon$ at either a fixed value of $R_Z$ or $R_\xi$. Using $\bar{U}_6$
instead of $\bar{U}_4$ leads to very similar results.  $\bar{R}_\xi$ at a fixed
value of $R_Z$ is little useful, because the corrections to scaling
and the statistical errors are relatively large.  In order
to get an idea of the corrections to scaling, we give results for the
fit intervals $5\le L \le 12$ and $10\le L \le 24$.
The difference of the two results
should be a rough estimate of the error due to next-to-leading
corrections to scaling.  As our final result we quote
\begin{equation}
\omega=0.785(20),
\label{omegaest}
\end{equation}
which includes (almost) all results for the interval
$10\le L \le 24$.

\subsection{Determination of $\lambda^*$ and $D^*$}
\label{ladstar}

\begin{figure}[tp]
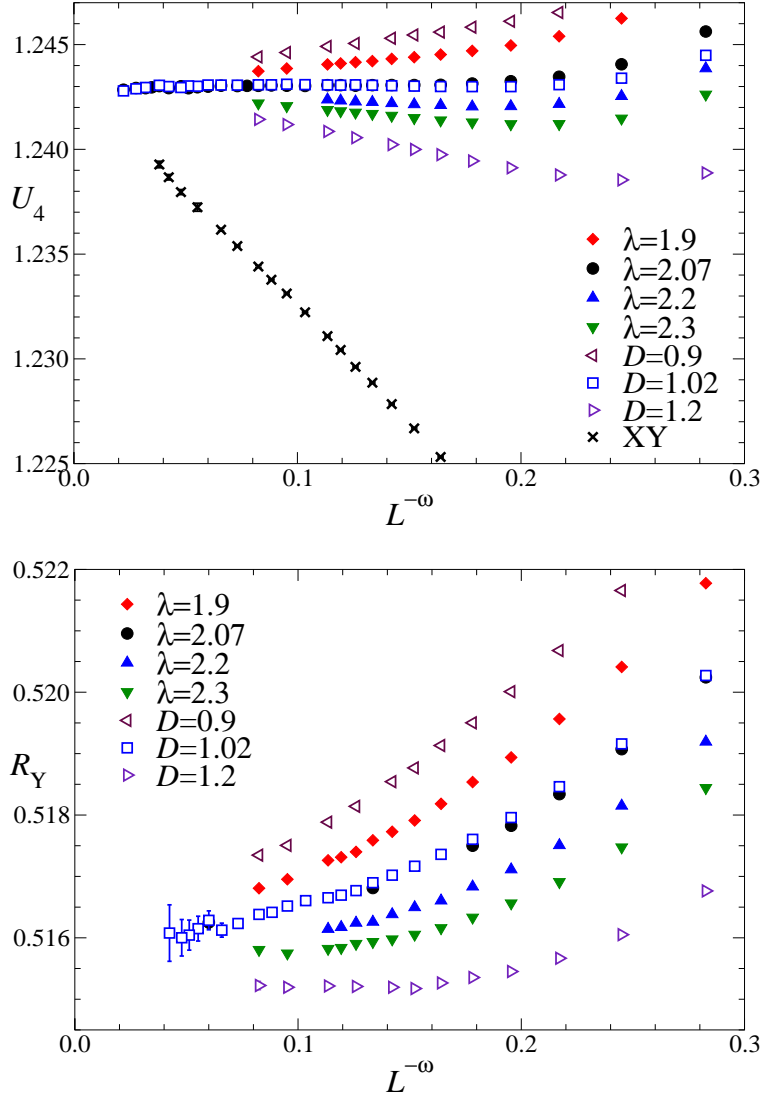

\centerline{\psfig{width=10truecm,angle=0,file=birzome.eps}}
\vskip 4mm
\centerline{\psfig{width=10truecm,angle=0,file=heliome.eps}}
\vskip 2mm
\caption{\label{omeevi}
$\bar{U}_4$ (above) and $\bar{R}_\Upsilon$ (below)
at fixed $R_Z=0.3202$ for various values of $\lambda$
($\phi^4$ model),  $D$ (ddXY model) and the standard XY model,
vs.\ $L^{-\omega}$ with $\omega=0.785$.
}
\end{figure}

To begin with, in Fig.~\ref{omeevi} we show  results for $\bar{U}_4$ and
$\bar{R}_\Upsilon$ at fixed $R_Z=0.3202$, for various values of $\lambda$ in
the $\phi^4$ model, of $D$ in the ddXY model, and in the standard XY model,
vs.\ $L^{-\omega}$ with $\omega=0.785$.  They show a clear evidence
of the leading scaling corrections, and of the existence of optimal
values $\lambda^*$, $D^*$ of $\lambda$ and $D$ for which they are
suppressed.  We also note that $\bar{R}_\Upsilon$ is subject to larger
next-to-leading scaling corrections; we shall return to this point
later.

In order to determine $\lambda^*$ and $D^*$, we mainly use our data
generated for $\lambda=2.07$ and $D=1.02$.  We fit them using various
ans\"atze. The most simple one is
\begin{equation}
\label{fitcor0sub}
\bar{R} = \bar{R}^* +  c L^{-\omega},
\end{equation}
where $\bar{R}$ is defined in Eq.~(\ref{barr}).
Eq.~(\ref{fitcor0sub}) includes only leading corrections to scaling;
we fix $\omega=0.785$ as previously obtained.  We may also include
subleading corrections to scaling and consider
\begin{equation}
\label{fitcor1sub}
\bar{R} = \bar{R}^* +  c L^{-\omega} + e L^{-\omega_2}.
\end{equation}
These fits are a bit problematic, since there are several
scaling corrections that have similar exponents with 
$1.8 \lesssim \omega_n \lesssim 2.0$, see Sec.~\ref{FSS}.  
In our fits we use both $\omega_2 = 1.8$ or $\omega_2 = 2$.

We first perform fits of types (\ref{fitcor0sub}) and (\ref{fitcor1sub}) for
the two models $\phi^4$ and ddXY separately. As $\bar R$ we consider
$\bar{U}_4$ at fixed $R_Z$, $\bar{U}_4$ at fixed $R_\xi$, $\bar{U}_6$ at fixed
$R_Z$, $\bar{U}_6$ at fixed $R_\xi$, and $\bar{R}_\Upsilon$ at fixed $R_Z$.
The results for $\bar R^*$ are, as required by universality, consistent for
the two models.  Hence we take our final results from joint fits of the
results for both models. For instance, in a joint fit with
ansatz~(\ref{fitcor0sub}) there are three free parameters: $\bar R^*$ and a
correction-to-scaling amplitude for each of the $\phi^4$ and ddXY models.

In order to determine $\lambda^*$ (and analogously $D^*$) we assume 
$c(\lambda)$ to be linear in a neighborhood of  $\lambda^*$ and 
write 
\begin{equation}
c(\lambda) \approx \frak{c}_1 (\lambda - \lambda^*), 
\label{linapprox}
\end{equation}
so that 
\begin{equation}
 \lambda^* = \lambda -  {1\over \frak{c}_1} c(\lambda).
\end{equation}
We use $\lambda = 2.07$, the value for which we have most of the 
simulations, and determine $\frak{c}_1$ by using
\begin{eqnarray}
&& \frak{c}_1 = 
 \left. {\partial c\over \partial \lambda}\right|_{\lambda=\lambda^*}
\approx \frac{c(\lambda=2.3) - c(\lambda=1.9)}{2.3-1.9}. \label{findiff} 
\end{eqnarray}
In the ddXY model we use the same formulas with $D = 1.02$ and 
\begin{eqnarray}
&&\frak{c}_1 = \left. {\partial c\over \partial D}\right|_{D = D^*}
\approx \frac{c(D=1.2) - c(D=0.9)}{1.2-0.9}. \label{findiff2}
\end{eqnarray}
In order to determine the needed values of $c(\lambda)$ and $c(D)$, 
we fix  $\omega=0.785$ (and, to estimate errors, $\omega=0.765$,
$\omega=0.805$). The results of the fits with ansatz~(\ref{correctionfit}) 
for $\omega=0.785$ are summarized in
Table \ref{tabcorr2}. The final estimate is taken from the fits with
$12\le L\le 24$.
The comparison with the fits with $6\le L \le 12$
gives us an idea of the error due to subleading corrections. 
It is small enough to be ignored in the following.

\begin{table}
\squeezetable
\caption
{Fits to ansatz~(\ref{correctionfit}) as in Table~\ref{tabcorr1},
but keeping $\omega$ fixed. $\Delta c\equiv c(\lambda=2.3) -c(\lambda=1.9)$
for the $\phi^4$ model and $\Delta c\equiv c(D=1.2) -c(D=0.9)$
for the ddXY model.}
\label{tabcorr2}
\begin{ruledtabular}
\begin{tabular}{lcccrrlr}
\multicolumn{1}{c}{model}&
\multicolumn{1}{c}{$R_1$}&
\multicolumn{1}{c}{$R_2$}&
\multicolumn{1}{c}{$\omega$}&
\multicolumn{1}{c}{$L_{\rm min}$}&
\multicolumn{1}{c}{$L_{\rm max}$}&
\multicolumn{1}{c}{$\Delta c$}&
\multicolumn{1}{c}{$\chi^2/{\rm d.o.f.}$}\\
\colrule
$\phi^4$ & $U_4$ & $R_Z$ & 0.785 &  6   &   12 & $-$0.01925(3) & 5.8  \\
         & & & & 12   &   24 & $-$0.01898(5) & 1.5 \\
         & & & 0.805  &  6   &   12 & $-$0.02006(3) & 1.5 \\
         & & & & 12   &   24 & $-$0.02002(5) & 1.2 \\
         & & & 0.765  &  6   &   12 & $-$0.01847(3) & 13.6 \\
         & & & & 12   &   24 & $-$0.01799(5) &  2.4 \\

& $U_4$ & $R_\xi$ & 0.785 &  6   &   12 & $-$0.02145(3) & 0.2 \\
& & & & 12   &   24 & $-$0.02145(5) & 1.6 \\

& $U_6$ & $R_Z$ & 0.785  &  6   &   12 & $-$0.0688(1)  & 7.7 \\
& & & & 12   &   24 & $-$0.0678(2)  & 1.5 \\

& $U_6$ & $R_\xi$ & 0.785&  6   &   12 & $-$0.0763(1)  & 0.2 \\
& & & & 12   &   24 & $-$0.0763(2)  & 1.6 \\

& $R_\Upsilon$ & $R_Z$ &0.785&  6 &12 & $-$0.00608(2) & 2.5 \\
& & & & 12 &24 & $-$0.00615(5) & 0.8 \\

ddXY & $U_4$ & $R_Z$ &  0.785 & 6   &   12 &  $-$0.03576(3)  & 0.8 \\
     &  & & & 12 & 24 & $-$0.03578(8)  & 0.9 \\
     & & & 0.805  &  6   &   12 &  $-$0.03727(3)  & 6.5 \\
     & & &  & 12   &   24 &  $-$0.03774(8)  & 3.4 \\
     & & & 0.765  &  6   &   12 &  $-$0.03431(3)  & 4.4 \\
     & & & & 12   &   24 &  $-$0.03393(7)  & 0.2 \\
     & $U_4$ & $R_\xi$ &  0.785 & 6   &   12 &  $-$0.04021(4)  &12.7  \\
     & & & & 12   &   24 &  $-$0.04077(9)  & 1.0  \\
     & $U_6$ & $R_Z$ & 0.785 & 6   &   12 &  $-$0.1271(1)  & 1.7 \\
     & & & & 12   &   24 &  $-$0.1276(3)  & 1.2 \\
     & $U_6$ & $R_\xi$ & 0.785 & 6   &   12 &  $-$0.1423(1)  & 22.3\\
     & & & & 12   &   24 &  $-$0.1447(3)  &1.3 \\
     & $R_\Upsilon$ & $R_Z$& 0.785 & 6 &12 & $-$0.01157(2) & 7.2 \\
     & & & & 12 & 24 & $-$0.01177(8) & 1.2\\
\end{tabular}
\end{ruledtabular}
\end{table}

\begin{figure}[tp]
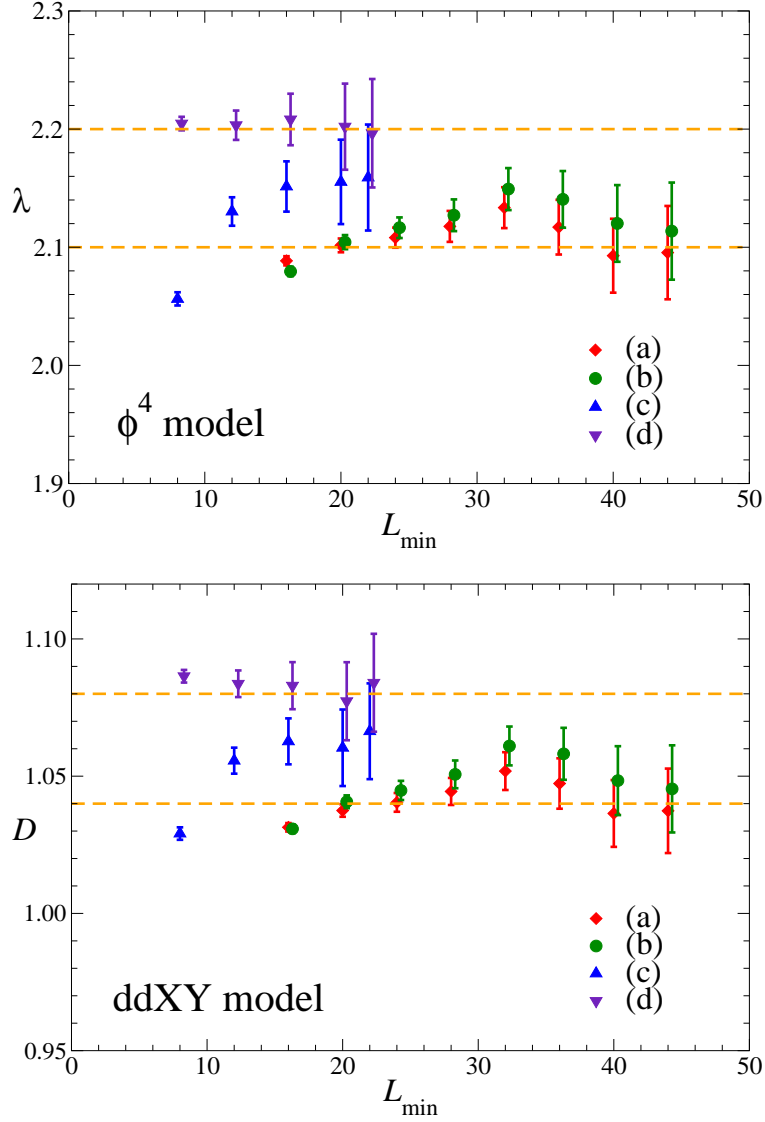

\centerline{\psfig{width=10truecm,angle=0,file=lambda.eps}}
\vskip 4mm
\centerline{\psfig{width=10truecm,angle=0,file=di.eps}}
\vskip 2mm
\caption{\label{ladistar}
Determination of $\lambda^*$ and $D^*$:  (a) results from fits of $\bar{U}_4$ at
$R_Z=0.3202$ to $a+c L^{-\omega}$ with $\omega=0.785$; (b) results from
fits of $\bar{U}_4$ at $R_\xi=0.5925$ to $a+c L^{-\omega}$ with $\omega=0.785$; (c)
results from fits of $\bar{U}_4$ at $R_Z=0.3202$ to $a+c L^{-\omega}+e
L^{-\omega_2}$ with $\omega=0.785$, $\omega_2=1.8$; (d) results from fits
of $\bar{U}_4$ at $R_\xi=0.5925$ to $a+c L^{-\omega}+e L^{-\omega_2}$ with
$\omega=0.785$, $\omega_2=1.8$.  The dashed lines indicate our
final estimates.
}
\end{figure}

We also checked whether the linear
approximation is sufficiently accurate by determining the derivatives
(\ref{findiff}) and (\ref{findiff2}) 
from other pairs of values of $\lambda$ and $D$.  The error
due to the linear extrapolation is approximately $10\%$, which is negligible
for the purpose of determining $\lambda^*$ and $D^*$. 
In Fig.~\ref{ladistar} we
plot the results for $\lambda^*$ and $D^*$ as functions of the $L_{\rm min}$
used in the fits to (\ref{fitcor0sub}) and (\ref{fitcor1sub}). 
The results of the fits to Eq.~(\ref{fitcor0sub}) show a systematic drift
and become stable only for $L_{\rm min} = 30$. This systematic variation
is mostly due to the next-to-leading corrections and indeed, fits 
to Eq.~(\ref{fitcor1sub}) are less dependent on $L_{\rm min}$ and give 
fully consistent results.  As our final result we quote
\begin{equation}
\lambda^*=2.15(5), \qquad D^*=1.06(2),
\label{ladstarest}
\end{equation}
which correspond to
\begin{equation}
(\bar{U}_4|_{R_Z=0.3202})^* =1.24281(10),\qquad
(\bar{U}_4|_{R_\xi=0.5925})^* =1.24277(10).
\end{equation}
Consistent results for $\lambda^*$ and $D^*$ are obtained by analyzing
$\bar{U}_6$ at $R_Z=0.3202$ and $\bar{R}_\xi=0.5925$.  Note that the estimates
of $\lambda^*$ and $D^*$ are slightly larger than those obtained in our
previous work,\cite{CHPRV-01} where we reported $\lambda^*=2.07(5)$ and
$D^*=1.02(2)$.  Since the larger statistics and larger lattice sizes allow us
to achieve a better control of all sources of systematic errors, and in
particular of the next-to-leading scaling corrections, we are confident that
the new estimates (\ref{ladstarest}) are now correct with the quoted errors
(that are as large as those reported in our previous work).

We also performed some MC simulations of the standard XY model
up to lattice size $L=96$.
Using the estimates for $\omega$ and $\bar{U}_4^*$ obtained above,
one-parameter fits of the MC data  give
(taking data for $24\le L\le 96$).
\begin{equation}
 \bar{U}_4|_{R_Z=0.3202} =1.24281 - 0.1014(4) L^{-0.785}
\end{equation}
and
\begin{equation}
 \bar{U}_4|_{R_\xi=0.5925} =1.24277 - 0.1138(4) L^{-0.785}\; .
\end{equation}
We can use these results to obtain a conservative  upper bound on the ratios
$|c(\lambda = 2.15)/c({\rm XY})|$ and 
$|c(D = 1.06)/c({\rm XY})|$ that are independent on the quantity one
is considering. Using the estimate of $\partial c/\partial \lambda|_{\lambda=\lambda^*}$
and taking into account that the error on $\lambda^*$ is $\pm 0.05$, we 
infer that the leading scaling-correction amplitude of
$U_4$ at $R_Z=0.3202$ for $\lambda = 2.15$ satisfies
$|c(\lambda = 2.15)| < 0.0024$.  The same bound is obtained in 
the ddXY model for $c(D=1.06)$. This implies 
$|c(\lambda=2.15)/c({\rm XY})| < 0.0024/0.1014 \approx 1/42$ and an 
analogous bound for $|c(D = 1.06)/c({\rm XY})|$. 
A similar calculation also shows that $|c(\lambda=2.07)|$ and 
$|c(D = 1.02)|$ are at least 20 times smaller than $|c({\rm XY})|$.

\subsection{Next-to-leading scaling corrections}
\label{omega2}

In this subsection we present evidence of next-to-leading scaling
corrections characterized by an exponent $\omega_{\rm nlo}\approx 2$,
as expected due to the presence of several irrelevant perturbations
with $y\approx -2$, as discussed in Sec.~\ref{FSS}.  In particular,
this provides a robust evidence of the absence of $1/L$ analytic
corrections, see discussion at the end of Sec.~\ref{FSS}.

We first construct improved variables $\bar{U}_4|_{R_Z=0.3202}$,
$\bar{U}_4|_{R_\xi=0.5925}$, and $\bar{R}_\Upsilon |_{R_Z=0.3202}$, 
which do not have leading scaling corrections. In the ddXY model 
(analogous formulas hold in the $\phi^4$ model by replacing $D$ with 
$\lambda$) we consider 
\begin{equation}
\bar{R}_{\rm imp} \approx \bar{R}(D_1)^{x} \bar{R}(D_2)^{1-x}.
\label{ttt}
\end{equation}
Expanding $\bar{R}$ as in Eq.~(\ref{fitcor0sub}) and using a linear 
approximation for $c(D)$, $c(D) \approx \frak{c}_1 (D - D^*)$,
as in the previous section, we obtain for $L\to \infty$
\begin{equation}
\bar{R}_{\rm imp} \approx 
    \bar{R}^*\left[1 + {\frak{c}_1\over \bar{R}^* L^\omega} 
   (x D_1 + (1 - x) D_2 - D^*) + \ldots \right].
\end{equation}
Thus, if we take $x = x^* = (D^*-D_2)/(D_1 - D_2)$, 
the leading scaling correction
cancels. In the ddXY model we use
the data at $D_1 = 1.02$ and $D_2 = 1.2$, while in the $\phi^4$ model
we combine our data for $\lambda_1=1.9$ and $\lambda_2=2.3$.

\begin{figure}[tp]
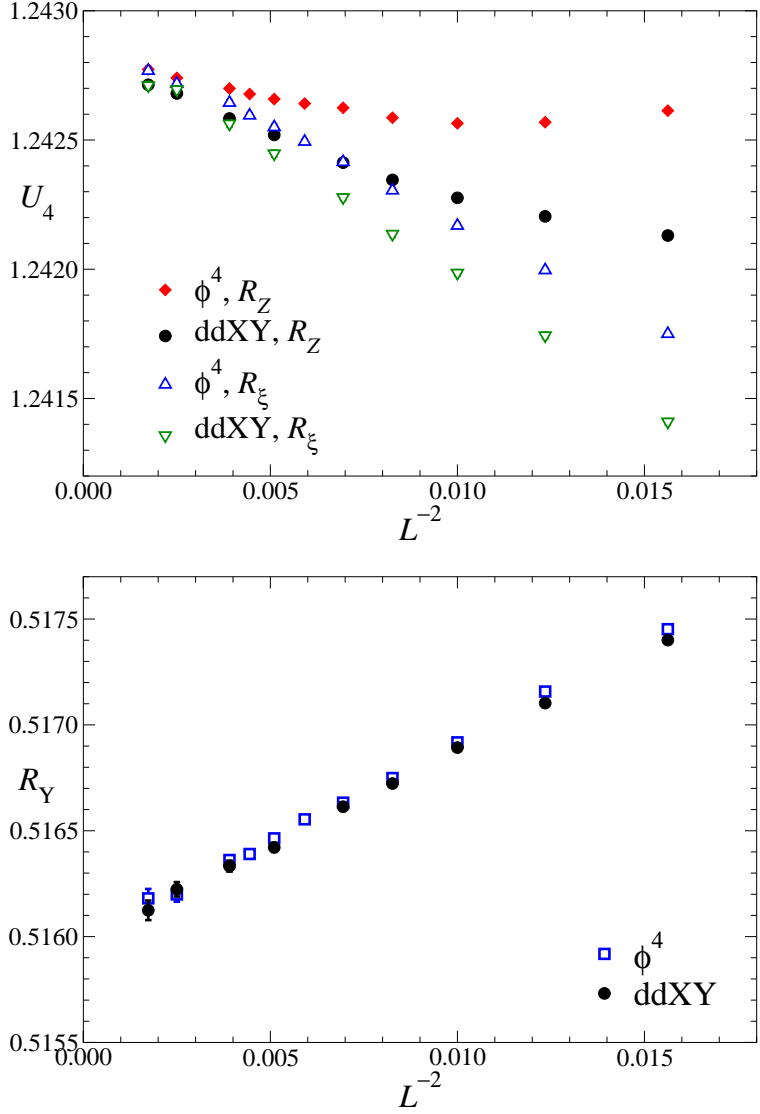

\centerline{\psfig{width=10truecm,angle=0,file=bitwo.eps}}
\vskip 4mm
\centerline{\psfig{width=10truecm,angle=0,file=helitwo.eps}}
\vskip 2mm
\caption{\label{ome2evi}
Estimates of the improved quantity (\ref{ttt}) in the case
of the Binder cumulant $U_4$ at $R_Z=0.3202$ and $R_\xi=0.5925$ (above),
and of $R_\Upsilon$ at $R_Z=0.3202$ (below),
vs.\ $L^{-2}$.
}
\end{figure}

In Fig.~\ref{ome2evi} we show  the improved combination (\ref{ttt}) in the case
of the Binder cumulant $\bar{U}_4$ as a function of $L^{-2}$.  For
$\bar{U}_4|_{R_\xi=0.5925}$ we observe straight lines to a good approximation.
On the other hand, for $\bar{U}_4|_{R_Z=0.3202}$ there is a clear bending of
the curves, indicating that, in the given range of lattice sizes,
corrections with exponent $\omega^\prime > 2$ are contributing
significantly.  We should note that these results do not provide a
completely independent information, since we have already used
$\omega_2 \approx 2$ as input in the determination of $D^*$ and
$\lambda^*$ in the previous section. Therefore, they only provide a
consistency check.  This is not the case for 
$R_\Upsilon\equiv \Upsilon L$, since
it was not used for the determination of $D^*$ and $\lambda^*$.  In
Fig.~\ref{ome2evi}, $\bar{R}_\Upsilon |_{R_Z=0.3202}$ is
plotted as a function of $L^{-2}$. We clearly observe a straight line. Hence
corrections with $\omega_2\approx 2$
clearly dominate in the whole range of lattice sizes that are
shown.  There are no leading
corrections to scaling and, as expected from general RG arguments,
also no corrections proportional to
$L^{-1}$. If there were a correction proportional to $L^{-1}$, the
ratio of its amplitude with the amplitude of the correction proportional
to $L^{-\omega}$ (i.e., the leading one) would not be the
same in different quantities. In other words, if corrections
proportional to $L^{-1}$ and $L^{-\omega}$ effectively cancel
for $U_4$ at our estimates of $D^*,\lambda^*$ and
for the range of values of $L$ considered,
there is no reason why this should also happen for $R_\Upsilon$.
Hence we conclude that our numerical results confirm the theoretical
argument that no $L^{-1}$ corrections
are present in FSS for periodic boundary conditions.

\subsection{The critical exponent $\nu$}
\label{nuest}

Here, we compute the critical exponent $\nu$ from a FSS analysis
of  the derivative $S_1\equiv \partial R_1/ \partial\beta$
at a fixed value of another (or the same) phenomenological coupling $R_2$.

\begin{table}
\squeezetable
\caption
{ Fits to ansatz~(\ref{deltas}) of $S_{U_4}$ at
$R_Z=0.3202$ for the ddXY model, where we have used $D_1=0.9$ and
$D_2=1.2$.  We fix $\omega=0.785$. $L_{\rm min}$ and $L_{\rm max}$ are the
minimal and maximal lattice size included in the fit.
}
\label{tabcorrnu}
\begin{ruledtabular}
\begin{tabular}{rcccc}
\multicolumn{1}{c}{$L_{\rm min}$}&
\multicolumn{1}{c}{$L_{\rm max}$}&
\multicolumn{1}{c}{$a(D=1.2)/a(D=0.9)$}&
\multicolumn{1}{c}{$c(D=1.2)-c(D=0.9)$}&
\multicolumn{1}{c}{$\chi^2$/d.o.f.}\\
\colrule
   6 & 12  &  0.9811(4)  & $-$0.0735(18)  &  1.4 \\
   6 & 24  &  0.9809(2)  & $-$0.0723(12)  &  1.0 \\
   8 & 24  &  0.9806(3)  & $-$0.0699(21)  &  0.9 \\
  10 & 24  &  0.9803(5)  & $-$0.0677(33)  &  1.1 \\
  12 & 24  &  0.9803(7)  & $-$0.0672(54)  &  0.3 \\
\end{tabular}
\end{ruledtabular}
\end{table}

For this purpose we define an improved quantity that does not 
have leading scaling corrections. Since $S_1$ behaves as 
\begin{equation}
S_1=a(\lambda) L^{1/\nu} \left[1 + c(\lambda)  L^{-\omega} + \ldots \right]
\end{equation}
for $L\to \infty$, if we take $\lambda\approx \lambda^*$, so that 
$c(\lambda)$ is small and the linear approximation 
$c(\lambda)\approx \frak{c}_1 (\lambda-\lambda^*)$
works well, an improved 
variable is simply
\begin{equation}
\label{shiftslope}
S_{1,\rm imp}(\lambda) \approx S_1(\lambda)
\left[ 1 - \frak{c}_1
(\lambda-\lambda^*)  L^{-\omega} \right].
\end{equation}
We compute $\frak{c}_1$ using
\begin{equation}
   \frak{c}_1 = 
   \left . {\partial c\over \partial \lambda} \right |_{\lambda=\lambda^*} \approx
    {c(\lambda_1) - c(\lambda_2) \over \lambda_1 - \lambda_2},
\end{equation}
where
$\lambda_1$ and $\lambda_2$ are sufficiently close to $\lambda^*$ 
so that the linear approximation works well. We estimate the difference 
$c(\lambda_1) - c(\lambda_2)$ from fits of 
\begin{equation}
\label{deltas}
\Delta S_1(\lambda_1,\lambda_2)\equiv
{ S_1(\lambda_2)|_{R_2={\rm const}} \over S_1(\lambda_1)|_{R_2={\rm const}} }
  = \frac{a(\lambda_2)}{a(\lambda_1)} \;
   \left\{ 1 + \left[ c(\lambda_2)-c(\lambda_1)\right]
    L^{-\omega} + \ldots \right\}.
\end{equation}
In the $\phi^4$ model we take $\lambda_1 = 1.9$, 
$\lambda_2 = 2.3$, and $\lambda = 2.07$ and fix $\omega = 0.785$.
The same formulas hold in the ddXY model: we take 
$D_1 = 0.9$, $D_2 = 1.2$, $D = 1.02$.

As an example,
let us discuss the determination of 
$c(D=0.9) - c(D = 1.2)$ for $S_{U_4}$ at $R_Z=0.3202$. In Table
\ref{tabcorrnu} we report some results of fits with
ansatz~(\ref{deltas}). They are quite
stable when $L_{\rm min}$ and $L_{\rm max}$ are varied.  This
indicates that leading scaling corrections dominate in the
difference and that subleading corrections vary little with $D$.
As our final result we take
\begin{equation}
c(D=1.2)-c(D=0.9)=-0.067(7),
\label{cestimate}
\end{equation}
which should take into account both
statistical and systematic errors.  The error due to the 
uncertainty on $\omega$ as well as 
the error due to the linear approximation are negligible. We
repeat this procedure for all quantities of interest.  Typically,
the amplitude differences analogous to (\ref{cestimate})
can be determined with an error of approximately $10\%$ for
the ddXY model and $15\%$ for the $\phi^4$ model.

To determine $\nu$, we fit the data of 
$S_1 = S_{U_4}$ at $R_2 = R_Z=0.3202$ to 
\begin{equation}
\label{nufit}
\left . S_1 \right |_{R_2={\rm const}}
\equiv
\left . \frac{\partial{R_1}}{\partial{\beta}}
\right |_{R_2={\rm const}} = a L^{1/\nu},
\end{equation}
and
\begin{equation}
\label{nufit2}
\left . S_1  \right |_{R_2={\rm const}}
  = a L^{1/\nu} \left(1 + e L^{-\omega_2} \right)
\end{equation}
where we consider either $\omega_2=1.8$ or $\omega_2=2$. 
Assuming that there are no leading scaling corrections,
the fits of the original data at $\lambda=2.07$ and $D=1.02$ 
give the result $\nu = 0.67181(12)$ and $\nu=0.67195(13)$
for $\lambda=2.07$ and $D=1.02$, respectively.
The errors take into account the results obtained by using the 
two ans\"atze and the $L_{\rm min}$ dependence.

In order to evaluate the effect of the residual leading corrections to
scaling, we repeat the same fits for $S_{1,\rm imp}$ computed by using
Eq.~(\ref{shiftslope}). We obtain estimates of $\nu$ that are smaller by 
roughly $0.0002$.  This change depends slightly on the ansatz and 
$L_{\rm min}$. These results allow us to give an effective estimate
of $\nu$ as a function of $D$ and $\lambda$, the $\lambda$ and $D$ 
dependence being due to the residual leading corrections to
scaling that are not taken into account in the fit. Since 
the estimates of the improved quantities correspond approximately
to those that would be obtained by using data at $\lambda = \lambda^*$ 
or $D = D^*$, we obtain
\begin{eqnarray}
&&\nu = 0.67181(12) - 0.0022(3)\times (\lambda-2.07) \quad
    \hbox{for the $\phi^4$ model}, \label{fssla} \\
&&\nu = 0.67195(13) - 0.0061(9)\times (D-1.02)\quad
    \hbox{for the ddXY model}.
\label{fssdi}
\end{eqnarray}
The $\lambda$ and $D$ dependence is that corresponding to the fit
without corrections with $L_{\rm min}=16$.
The error on the linear coefficient is due to the error on
${\partial c/\partial \lambda}$.
Using the estimates $\lambda^*=2.15(5)$ and
$D^*=1.06(2)$, we obtain the results
\begin{eqnarray}
&&\nu = 0.67163(12)[11] \quad \hbox{for the $\phi^4$ model},
\label{nuestphi4a}\\
&&\nu = 0.67171(13)[12] \quad \hbox{for the ddXY model},
\label{nuestddxya}
\end{eqnarray}
where the first error is statistical and the second one is due to the
uncertainty on $\lambda^*$ and $D^*$.
The analysis of
$S_{U_4}$ at $R_\xi=0.5925$ gives analogous results.
Other quantities provide consistent, but less precise, estimates.

Finally, we analyze the derivative $S_{R_\Upsilon}\equiv \partial R_\Upsilon
/\partial \beta$ of the helicity modulus. In this case subleading corrections
are smaller when considering the data at $\beta = \beta_c$, rather than at a
fixed phenomenological coupling. Since we have little data for the $\phi^4$
model at $\lambda=2.07$, we only give results for the ddXY model. We first
compute the improved slopes $S_{R_\Upsilon,\rm imp}$ by using
Eq.~(\ref{shiftslope}).  Fits to ansatz $S = a L^{1/\nu}$ have a small
$\chi^2/{\rm d.o.f.}$ (d.o.f. is the number of degrees of freedom of the fit)
already from rather small $L_{\rm min}$, indicating that subleading
corrections are rather small in this quantity.  Fitting the data with $16\le L
\le 128$, we obtain $\nu=0.67200(15)$ at $D=1.02$ and $\nu=0.67173(15)$ at
$D=1.06$ (more precisely, for the improved slope).  Therefore, taking also
into account the error on $D^*$, we might quote $\nu=0.6717(3)$ as final
result of this analysis, which agrees with the results (\ref{nuestphi4a}) and
(\ref{nuestddxya}).  We have also checked the dependence of this result on the
estimate of $\beta_c$, finding that the error due the uncertainty of $\beta_c$
is definitely smaller than the error on $\nu$ quoted above.

\subsection{Eliminating leading scaling corrections from the derivative of
phenomenological couplings}
\label{ellea}

We now consider a combination of the derivative 
$S_i\equiv \partial R_i/\partial \beta$
of two phenomenological couplings $R_i$,
at $\beta_c$ or at fixed $R_j$,
\begin{eqnarray}
\left |S_1(\lambda) \right|^p \; \left |S_2(\lambda) \right|^{1-p},
\end{eqnarray}
and show that one can choose a value $p$ such that 
this quantity is improved---no leading
corrections to scaling---for any $\lambda$ or $D$. 
Moreover, the computation of $p$ does not rely on any
estimate of $\omega$.

For $L\to \infty$, $S_i$ at $\beta_c$ or at a fixed $R_j$, behaves as
\begin{equation}
S_i(\lambda) = a_i(\lambda) L^{1/\nu}
    \left[ 1 + c_i(\lambda) L^{-\omega} + \ldots \right]\; .
\end{equation}
Therefore, we have
\begin{eqnarray}
\left |S_1(\lambda) \right|^p \; \left |S_2(\lambda) \right|^{1-p}
= |a_1(\lambda)|^p \; |a_2(\lambda)|^{1-p} \; L^{1/\nu} \;
  \left\{ 1 +  \left[ p c_1(\lambda) + (1-p) c_2(\lambda) \right]
L^{-\omega} + \ldots \right\}.
\label{Scombination}
\end{eqnarray}
An improved quantity is obtained by taking $p = p^*$, where 
\begin{equation}
p^* c_1(\lambda) + (1-p^*) c_2(\lambda)=0.
\end{equation}
Note that, since the ratios
$c_1(\lambda)/c_2(\lambda)$ are universal, i.e., independent of $\lambda$ (or
$D$), the optimal value $p^*$ is universal.

Now we show how $p^*$ can be accurately computed.
We consider ratios of $S_i(\lambda)$ at different values of $\lambda$:
\begin{equation}
\label{diffcorrslope}
 \frac{S_i(\lambda_2)}{S_i(\lambda_1)} =
 \frac{a_i(\lambda_2)}{a_i(\lambda_1)}
   \left\{1 + \left[c_i(\lambda_2) - c_i(\lambda_1) \right] L^{-\omega} 
   + \ldots
\right\}\; .
\end{equation}
Due to the universality of the amplitude ratio we have
\begin{equation}
 \frac{c_2(\lambda_1)}{c_1(\lambda_1)} =
 \frac{c_2(\lambda_2)}{c_1(\lambda_2)} =
 \frac{c_2(\lambda_2)-c_2(\lambda_1)}{c_1(\lambda_2)-c_1(\lambda_1)}\; .
\end{equation}
Therefore,
\begin{eqnarray}
\left[\frac{S_1(\lambda_2)}{S_1(\lambda_1)} \right]^p
 \left[\frac{S_2(\lambda_2)}{S_2(\lambda_1)} \right]^{1-p} &=&
\left[\frac{a_1(\lambda_2)}{a_1(\lambda_1)}\right]^p
\left[\frac{a_1(\lambda_2)}{a_1(\lambda_1)}\right]^{1-p}
\\[3mm]
&& \times 
\left\{ 1 + \left[ p \left( c_1(\lambda_2) - c_1(\lambda_1) \right) + (1-p)
 \left( c_2(\lambda_2) - c_2(\lambda_1) \right) \right] 
    L^{-\omega} + \ldots   \right\}
\nonumber 
\end{eqnarray}
We can obtain the desired value of $p^*$ by imposing that the combination
\begin{equation}
\label{optifit}
 \left[\frac{S_1(\lambda_2)}{S_1(\lambda_1)} \right]^p
 \left[\frac{S_2(\lambda_2)}{S_2(\lambda_1)} \right]^{1-p} 
\end{equation}
is $L$ independent. 
This procedure does not require any knowledge of $\omega$ and
only assumes that leading scaling corrections dominate in the considered range of 
lattice sizes.

\begin{table}
\squeezetable
\caption
{ Fits to ansatz~(\ref{optifit}) of the improved combination of 
$S_{U_4}$ and $S_{R_Z}$ at $R_Z=0.3202$.
$L_{\rm min}$ and $L_{\rm max}$ are the
minimal and maximal lattice size included in the fit. For a detailed
discussion, see the text.
}
\label{tabcorrp}
\begin{ruledtabular}
\begin{tabular}{crrll}
\multicolumn{1}{c}{model}&
\multicolumn{1}{c}{$L_{\rm min}$}&
\multicolumn{1}{c}{$L_{\rm max}$}&
\multicolumn{1}{c}{$p$}&
\multicolumn{1}{c}{$\chi^2$/d.o.f.}\\
\colrule
%
$\phi^4$ & 6 & 12 & 
   0.703(11) &  0.93 \\
& 8 & 24 & 
   0.718(11) & 1.37 \\
& 12 & 24 & 
  0.744(30)  & 1.61 \\
%
%
 ddXY & 6 & 12 & 
 0.736(6) &  1.87  \\
& 8 & 24 & 
 0.720(8) &  1.29  \\
& 12 & 24 & 
 0.702(19)&  0.02  \\
\end{tabular}
\end{ruledtabular}
\end{table}

The optimal pair of slopes $S_i$ turns out to be
\begin{equation}
\left . S_1= {\partial R_Z \over \partial \beta} \right |_{R_Z=0.3202},
\qquad
\left . S_2= {\partial U_4 \over \partial \beta} \right |_{R_Z=0.3202},
\end{equation}
essentially because the amplitudes of their leading scaling correction
have opposite signs. In order to determine the corresponding value of
$p^*$, we perform fits with ansatz~(\ref{optifit}) using the data
at $\lambda_1=1.9$ and $\lambda_2=2.3$ for the $\phi^4$ model, and at
$D_1=0.9$ and $D_2=1.2$ for the ddXY model. Some results are reported
in Table~\ref{tabcorrp}. The estimates of $p^*$ from the two models
are consistent, as required by universality.  We take $p^*=0.72(3)$ as
our final estimate (from fits with $L_{\rm min}=8$, while the error is
estimated by varying the fit range).

\begin{figure}[tp]
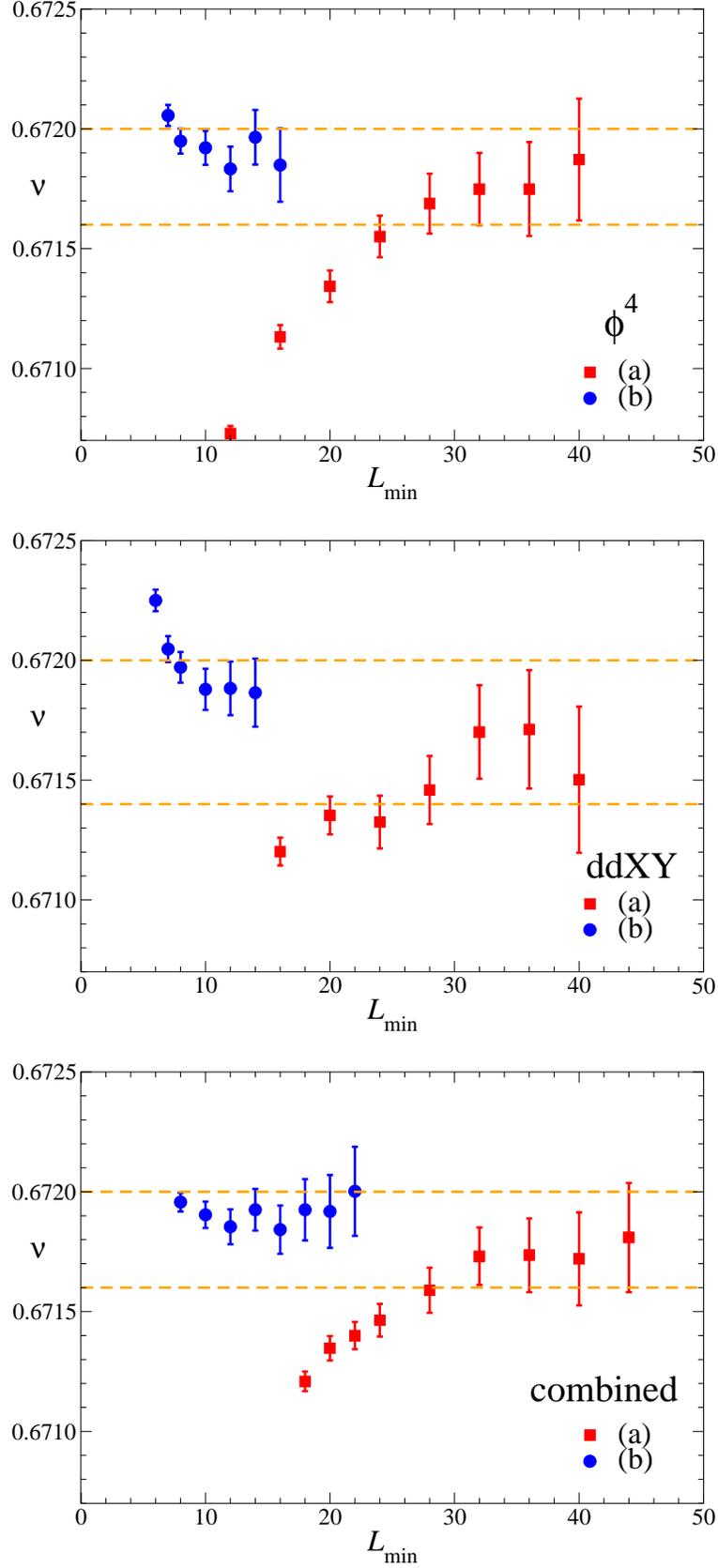

\centerline{\psfig{width=10truecm,angle=0,file=nuphi4FSS.eps}}
\vskip 4mm
\centerline{\psfig{width=10truecm,angle=0,file=nuddFSS.eps}}
\vskip 4mm
\centerline{\psfig{width=10truecm,angle=0,file=nucombFSS.eps}}
\vskip 2mm
\caption{\label{nuFSS}
Results for the exponent $\nu$ obtained by fits for several
values of $L_{\rm min}$: (a) to $a L^{1/\nu}$ and
(b) to $a L^{1/\nu}( 1 + e L^{-\omega_2})$ with $\omega_2=1.8$. }
\end{figure}

Using this estimate for $p^*$, we construct the improved
combinations (\ref{Scombination})
at $\lambda=2.07$ for the $\phi^4$ model and at
$D=1.02$ for the ddXY model. Since these values of $\lambda$ and $D$ are
close to the optimal values $\lambda^*$ and $D^*$, leading scaling
corrections are quite small. Therefore, the uncertainty on $p^*$
is negligible with respect to the final error of
our estimate for $\nu$.
In Fig.~\ref{nuFSS} we show results for the exponent $\nu$ obtained by
fits to the functions $a L^{1/\nu}$ and to $a L^{1/\nu}( 1 + e
L^{-\omega_2})$ with $\omega_2=1.8$, for several values of $L_{\rm min}$.
Guided by Fig.~\ref{nuFSS}, we take
\begin{eqnarray}
&&\nu = 0.6718(2) \quad \hbox{for the $\phi^4$ model},
\label{nuestphi4b}\\
&&\nu = 0.6717(3) \quad \hbox{for the ddXY model}
\label{nuestddxyb}
\end{eqnarray}
as our final estimates. Moreover,
combined fits applied to both $\phi^4$ and ddXY models
give the estimate $\nu=0.6718(2)$.

\begin{figure}[tp]
\centerline{\psfig{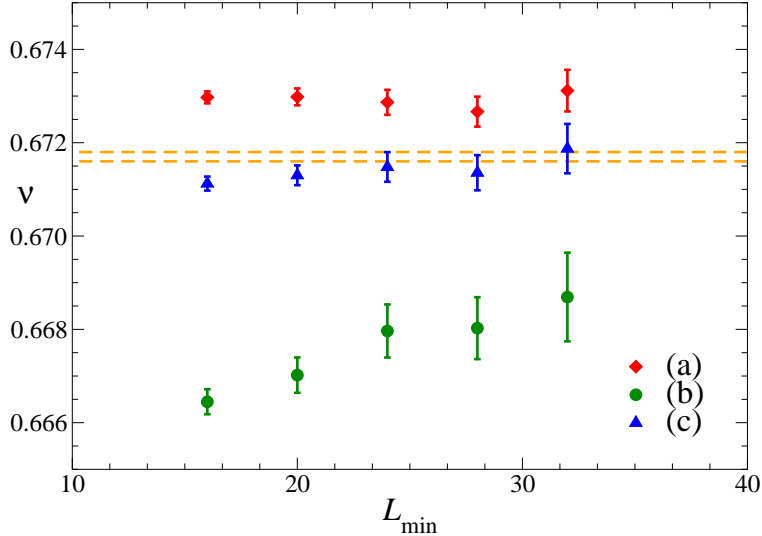}}
\vskip 2mm
\caption{\label{nustXY}
Results for the critical exponent $\nu$ from MC simulations
of the standard XY model, as obtained by fits ot the data of
the derivative of $R_Z$  at $R_Z=0.3202$ (a),
the derivative of $U_4$  at $R_Z=0.3202$ (b),
and by using the method that combines them to eliminate the residual
leading scaling corrections (c). The two dashed lines correspond to our
final estimate $\nu = 0.6717(1)$.
}
\end{figure}

As further check, we apply this method to the standard XY model,
where leading corrections to scaling are large.  In Fig.~\ref{nustXY}
we show results for the critical exponent $\nu$ from MC simulations of the
standard XY model up to $L=96$, as obtained by fits to the simple ansatz
$S=a L^{1/\nu}$ of the data of $S_{R_Z}$ at $R_Z=0.3202$, $S_{U_4}$ at
$R_Z=0.3202$, and their combination (\ref{Scombination}) at
$p=p^*=0.72$.  The first two sets of results clearly disagree
with the estimate of $\nu$ from improved Hamiltonians and also
between themselves (thus, they are inconsistent). Instead, the 
analysis of their improved combination
provides perfectly consistent results, giving further support
for the validity of the method and confirming the accurate determination of 
$p^*$.

\subsection{The critical exponent $\nu$ from the finite-size scaling
 of the energy density}
\label{energysec}

\begin{figure}[tp]
\centerline{\psfig{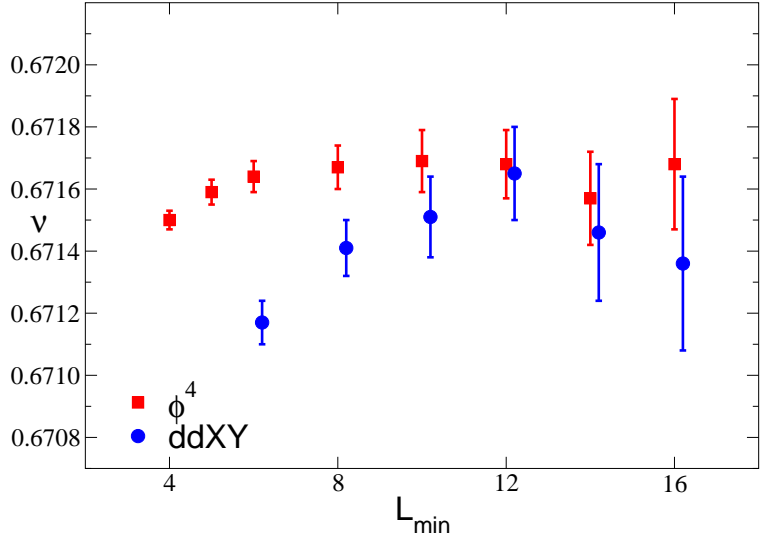}}
\vskip 2mm
\caption{\label{nueFSS}
Results of fits for the FSS of the energy for the $\phi^4$ model
at $\lambda=2.07$ and the ddXY model at $D=1.02$.
}
\end{figure}

We also derive estimates of $\nu$ from the FSS of the energy density.
On finite lattices of size $L^3$, the free energy density behaves as
\begin{equation}
 f(\beta,L)  = f_{ns}(\beta) + f_s(\beta,L),
\end{equation}
where the nonsingular part of the free energy density $f_{ns}$ does not
depend on $L$, apart from exponentially small contributions.
The singular part is expected to behave as
\begin{equation}
 f_s(\beta,L) =  L^{-d} g(t L^{1/\nu},u_3 L^{-\omega},\ldots ).
\end{equation}
The energy density is obtained by taking the derivative
with respect to $\beta$,
\begin{equation}
E(\beta,L) = E_{ns}(\beta) + 
   L^{-d+1/\nu} g_e(t L^{1/\nu},u_3 L^{-\omega},\ldots) + \ldots
\end{equation}
Setting $\beta=\beta_c$, we obtain for $L\to \infty$ the expansion
\begin{equation}
E(\beta_c,L)  = E_{ns}(\beta_c) + a L^{-d+1/\nu} (1+c L^{-\omega}) + 
  \ldots 
\label{esc}
\end{equation}
In Fig.~\ref{nueFSS} we show the results of the fits to Eq.~(\ref{esc})
without correction terms (i.e., with $c=0$)
for the $\phi^4$ and ddXY models respectively at
$\lambda=2.07$ and $D=1.02$.  Our final estimates of $\nu$ from the
scaling of the energy density
(obtained from fits of data for $12\le L\le 128$) are
\begin{eqnarray}
&&\nu = 0.6717(2) \quad \hbox{for the $\phi^4$ model},
\label{nuestphi4c}\\
&&\nu = 0.6715(3) \quad \hbox{for the ddXY model}.
\label{nuestddxy}
\end{eqnarray}
Our data at $\lambda=1.9,2.3$ in the case of the $\phi^4$ model and
$D=0.9,1.2$ for the ddXY model are not sufficient to get a reliable estimate
of the systematic error due to the leading scaling corrections.
However, we can estimate it by a comparison with the
standard XY model.  Fitting the data of the standard XY model to the simplest
ansatz without scaling corrections (using the value of $\beta_c$
reported in Table \ref{betac} and  data for
$8\le L \le 96$), we obtain $\nu=0.6701(2)$, i.e., the exponent is
underestimated by approximately $0.0015$. We may
use this difference to estimate the leading scaling corrections amplitude in
the standard XY model. Taking into account that the largest lattice for the 
XY model has $L=96$ instead of $L=128$, we obtain $c\approx 0.0015 \times
(128/96)^{-0.785} = 0.0012$, 
where we have also taken into account the difference
of the sizes $L$ used in the fits.
Since the leading correction amplitudes at $D=1.02$ and $\lambda=2.07$ should
be smaller by a factor of approximately 20 than in the standard XY model, we
may have a shift by $0.00006$ in our estimates (\ref{nuestphi4c}) and
(\ref{nuestddxy}) for $\nu$ due to leading corrections, which is much smaller
than their errors.

Note that, in the case of the energy, analyses at fixed
phenomenological coupling do not work. Since
\begin{equation}
 R(L,\beta) = R^* + a (\beta-\beta_c) L^{1/\nu} + b L^{-\omega} + \ldots, 
\end{equation}
and therefore
\begin{equation}
R_{f} = R^* + a (\beta_{f} -\beta_c) L^{1/\nu} + b L^{-\omega} + \ldots,
\end{equation}
then
\begin{equation}
\beta_{f} = \beta_c + a^{-1} (R_{f}-R^*)  L^{-1/\nu}
       - (b/a) L^{-1/\nu -\omega} + \ldots
\end{equation}
Plugging $\beta_{f}$ in the nonsingular part of the energy,  one obtains 
a term $O(L^{-1/\nu})$, which, in this particular case,
completely masks the interesting scaling term, since $1/\nu -3\approx
-1.51$ and $-1/\nu\approx -1.49$.

\subsection{The critical exponent $\eta$}
\label{etaestim}

We compute the critical exponent $\eta$ from the FSS behavior of the magnetic
susceptibility $\chi$ either at
$R_Z=0.3202$ or at $R_\xi=0.5925$.
As in the analyses of the derivatives $S_R$, we first compute an improved 
quantity for $\chi$ that does not have leading scaling corrections. 
Since
\begin{equation}
\chi = a L^{2-\eta} \left( 1+ c L^{-\omega} + \ldots \right),
\end{equation}
close to the improved value $\lambda^*$ where $c$ is small and 
a linear approximation $c(\lambda) = \frak{c}_1 (\lambda - \lambda^*)$ 
suffices, we can take
\begin{equation}
\chi_{\rm imp} = \chi(\lambda) \left(1 - c(\lambda) L^{-\omega}\right)
   \approx \chi(\lambda) \left[1 - \frak{c}_1
   (\lambda - \lambda^*) L^{-\omega}\right].
\label{chifitcor2}
\end{equation}
To compute $\frak{c}_1$ we consider the ratio
\begin{equation}
 \frac{\left . \chi(\lambda_2) \right |_{R_Z=0.3202}}
      {\left . \chi(\lambda_1) \right |_{R_Z=0.3202}} \approx
     \frac{a(\lambda_2)}{a(\lambda_1)} \;
      \left\{ 1 + \left[ c(\lambda_2)-c(\lambda_1)\right]
         L^{-\omega} + \ldots \right\}
\label{eq78}
\end{equation}
with $\lambda_2=2.3$ and $\lambda_2=1.9$. Fits to Eq.~(\ref{eq78}) allow us to
estimate the difference $c(\lambda=2.3)-c(\lambda=1.9)$. Then
\begin{equation}
 \frak{c}_1 = \left. {\partial c\over \partial \lambda} \right|_{\lambda=\lambda^*}
\approx  \frac{c(\lambda=2.3)-c(\lambda=1.9)}{2.3-1.9}.
\end{equation}
Analogous equations can be written for the ddXY model and at fixed $R_\xi$.

\begin{figure}[tp]
\centerline{\psfig{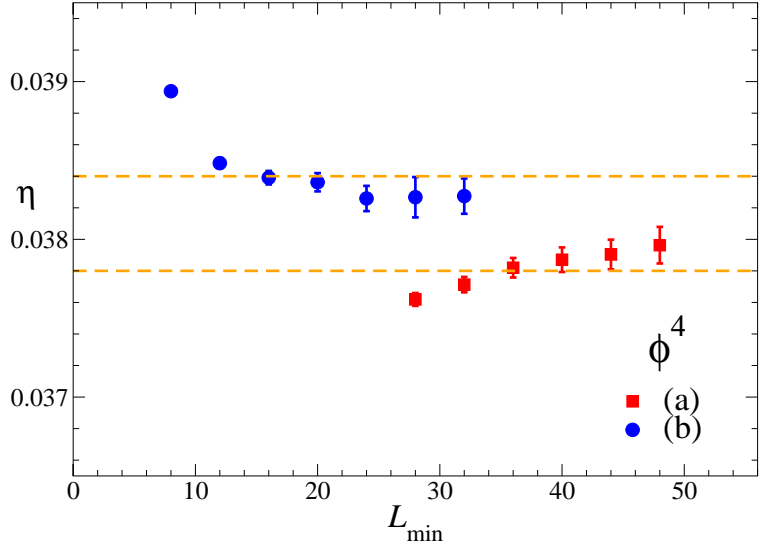}}
\vskip 2mm
\caption{\label{etaphi4FSS}
Results for the exponent $\eta$ obtained by fits of $\chi_{\rm imp}$
for several values of $L_{\rm min}$: (a) to $a L^{1/\nu}$ and
(b) to $a L^{1/\nu}( 1 + e L^{-\omega_2})$ with $\omega_2=1.8$.
The dashed lines indicate our final FSS estimate $\eta=0.0381(3)$.
}
\end{figure}

Then, we fit the data
for the improved $\chi$ [we use $\lambda = 2.07$ in Eq.~(\ref{chifitcor2})]
using the ans\"atze
\begin{eqnarray}
&&\left . \chi_{\rm imp} \right|_{R_Z=0.3202} = a L^{2-\eta}, \label{chifit} \\
&&\left . \chi_{\rm imp} \right|_{R_Z=0.3202} =
    a L^{2-\eta} \left( 1+ e L^{-\omega_2} \right),
\label{chifitcor}
\end{eqnarray}
where we fix either $\omega_2=1.8$ or $\omega_2=2$.  In Fig.~\ref{etaphi4FSS}
we show the results of the fits of $\chi_{\rm imp}|_{R_Z=0.3202}$ 
for the $\phi^4$ model
vs.\ the minimum value $L_{\rm min}$ of $L$ allowed in the fits.  Similar
results are obtained for the ddXY model, using $D = 1.02$ in 
Eq.~(\ref{chifitcor2}). In all cases, the fits with
ansatz (\ref{chifit}) have a large $\chi^2/{\rm d.o.f.}$ for $L_{\rm min}
\lesssim 40$. Moreover, the resulting values of $\eta$ appear to slightly
increase with increasing $L_{\rm min}$.  In contrast, the fits allowing for
subleading corrections are more stable and give a $\chi^2/{\rm d.o.f.}$ close
to 1 already for $L_{\rm min}\gtrsim 10$.  We also mention that fixing
$R_\xi$ instead of $R_Z$ gives slightly lower values for $\eta$, by about
0.00005 for the fit to (\ref{chifit}), and 0.00015 for the fits to
(\ref{chifitcor}).  A comparison of $\lambda=2.07$ and $\lambda=2.15$
indicates that a possible error due to the uncertainty of $\lambda^*$ should
be approximately 0.00005.  Our final FSS estimate obtained from both $\phi^4$ and
ddXY models is
\begin{equation}
\eta=0.0381(3).
\end{equation}

\section{Critical exponents from improved high-temperature expansions}
\label{HT}

\subsection{High-temperature expansions}
\label{series}

High-temperature (HT) expansion in powers of the inverse temperature $\beta$
is one of the most effective lattice approaches to investigate the critical
behavior in the HT phase.  We consider a general class of models defined on a
simple cubic lattice by the Hamiltonian
\begin{equation}
{\cal H} =  - \beta\sum_{\left<xy\right>} {\vec\phi}_x\cdot{\vec\phi}_y +\,
\sum_x V(\vec{\phi}^{\,2}),\label{hamiltonian}
\end{equation}
where $\beta\equiv 1/T$, $\left< xy \right>$ indicates nearest-neighbor sites,
$\vec{\phi}_x = (\phi_x^{(1)},\phi_x^{(2)})$ is a two-component real variable,
and $V(\vec{\phi}^{\,2})$ is a generic potential satisfying appropriate
stability constraints.

Using the linked-cluster expansion technique (see
Refs.~\onlinecite{Campostrini-01,CHPRV-01} for details), we extended the HT
computations of Ref.~\onlinecite{CHPRV-01} by adding a few terms in the HT
series.  The two main steps of the algorithm are the generation of the graphs
and the evaluation of the contribution of each graph.  The first step is
limited by memory, and was run on a computer with 16 Gbytes of RAM.  The
second step is limited by processing time; it was parallelized and required
approximately 4 years of CPU-time on a single 2.0 GHz Opteron processor.  We
computed the 22nd-order HT expansion of the magnetic susceptibility and the
second moment of the two-point function
\begin{eqnarray}
&&\chi = \sum_x \langle \phi_{\alpha}(0) \phi_{\alpha}(x) \rangle,
\label{chi}\\
&&m_2 = \sum_x x^2 \langle \phi_{\alpha}(0) \phi_{\alpha}(x) \rangle,
\nonumber
\end{eqnarray}
and therefore, of the second-moment correlation length
\begin{equation}
\xi^2 = {m_2 \over 6\chi}.
\end{equation}
Moreover, we computed the HT expansion of some zero-momentum connected
$2j$-point Green's functions $\chi_{2j}$
\begin{equation}
\chi_{2j} = \sum_{x_2,...,x_{2j}}
    \langle \phi_{\alpha_1}(0) \phi_{\alpha_1}(x_2) ...
        \phi_{\alpha_j}(x_{2j-1}) \phi_{\alpha_j}(x_{2j})\rangle_c
\label{chi2j}
\end{equation}
($\chi = \chi_2$). We computed $\chi_4$ to 20th order,
$\chi_6$ and $\chi_8$ to 18th order.

The HT series for the general model (\ref{hamiltonian}) will be reported
elsewhere.  In the following we will restrict ourselves to the $\phi^4$ and
ddXY models, cf.\ Eqs.\ (\ref{phi4Hamiltonian}) and (\ref{ddxy}).

\subsection{Critical exponents from the analysis of the HT expansion
            of improved models}
\label{resht}

In our analysis of the HT series we consider quasi-diagonal first-, second-,
and third-order integral approximants (IA1's, IA2's and IA3's respectively),
and in particular biased IA$n$'s (bIA$n$'s) using the most precise available
estimate of $\beta_c$.
The FSS estimates of $\beta_c$ are reported in Table~\ref{betac}.  We
refer to Ref.~\onlinecite{CHPRV-01}, and in particular to its Appendix
B, for details on the HT analysis and the precise definition of
the various integral approximants.  A review of methods for the
analysis of HT series can be found in Ref.~\onlinecite{Guttrev}.

The leading nonanalytic corrections are the dominant source of
systematic errors in HT studies.  Indeed, nonanalytic corrections
introduce large and dangerously undetectable systematic deviations in
the results of the analysis.  Integral approximants can in principle
cope with an asymptotic behavior of the form (\ref{chiwexp}); however,
in practice, they are not very effective when applied to the series of
moderate length available today.  As shown in
Refs.~\onlinecite{CHPRV-01,CPRV-02,CHPRV-02}, analyses of the HT series for
the improved models lead to a significant improvement in the estimates
of the critical exponents and of other infinite-volume HT quantities.
The crux of the method is a precise determination of the improved value
of the parameter appearing in the Hamiltonian. In this respect FSS
techniques appear quite effective, as
we have shown in the preceding section. A further improvement is achieved by
biasing the HT analysis using the available estimates of $\beta_c$.

Our working hypothesis is that, with the series of current length, the
systematic errors, i.e., the systematic deviations that are not taken
into account in the HT analysis, are largely due to the leading
nonanalytic corrections, especially when they are characterized by a
relatively small exponent, as is the case in the 3-$d$ XY universality
class where $\Delta=\omega\nu\approx 0.53$. Therefore, improved models are
expected to give results with smaller and, more importantly, reliable
error estimates. The systematic errors in our analyses are related either to 
next-to-leading nonanalytic scaling corrections or to our 
approximate knowledge of $\lambda^*$ and $D^*$. 
Since $\Delta_2=\omega_2 \nu \approx 1.2$, we will assume that 
next-to-leading corrections do not play much role, and 
we will only take into account the residual leading corrections
proportional to $\lambda-\lambda^*$ ($D-D^*$ for the ddXY model).
Of course, this 
hypothesis requires stringent checks.  A nontrivial check is achieved
by comparing the results obtained using different improved models: If
the hypothesis is correct, they should agree within error bars.

As an additional check of our results we compare IA1's
of the 22nd-order HT series of $\chi$ and $\xi^2$ with high-statistics MC
results.
We simulated the $\phi^4$ model at $\lambda=2.1$,
$2.2$ and the
dd-XY model  at $D=1.02$,  $1.2$ in the HT phase
($\beta<\beta_c$).
  We alternate single-cluster updates and local Metropolis
and overrelaxation updates.
In order to obtain negligible finite-size effects
(i.e., orders of magnitude smaller than the
statistical errors) we used lattices of  size $L > 10 \xi$
throughout. We obtained infinite-volume estimates up 
to $\xi \approx 30$ on a $350^3$ lattice.

\begin{figure}[tp]
\centerline{\psfig{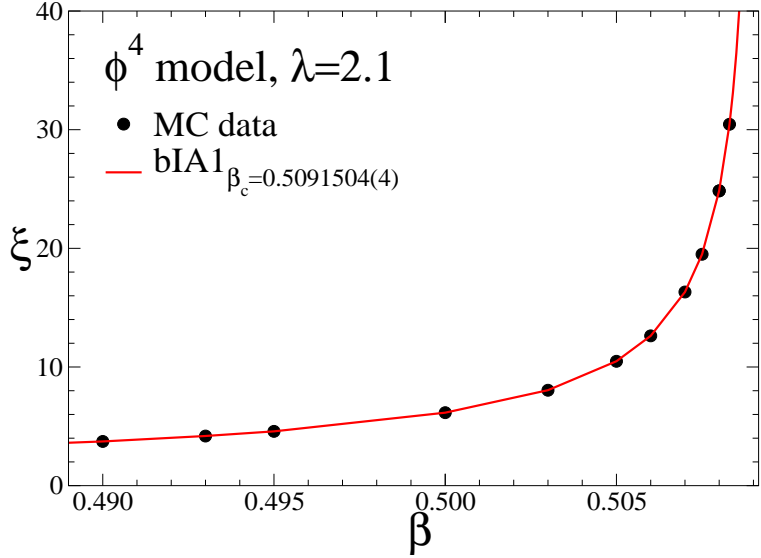}}
\vskip 2mm
\caption{\label{ximcht}
High-temperature MC data for the correlation length of the $\phi^4$
model at $\lambda=2.1$, and the result of the fit using bIA1 of the
22nd-order HT series of $\xi^2$.  }
\end{figure}

In Fig.~\ref{ximcht} we show MC data for
the $\phi^4$ model at $\lambda=2.1$ from $\beta=0.493$, where
$\xi=4.1825(2)$, to the largest $\beta$ value $\beta=0.5083$, where
$\xi=30.453(10)$.  bIA1's using the FSS estimate
$\beta_c=0.5091503(5)$ provide perfectly consistent results, for
example $\xi=30.449(1)[7]$, where the first error is related to the
spread of the approximants and the second one to the uncertainty on
$\beta_c$.  We also obtain an independent estimate of $\beta_c$ by
fitting the MC data of $\xi$ to bIA1's with $\beta_c$ taken as a free
parameter. The resulting estimate $\beta_c=0.5091504(4)$ (with
$\chi^2/{\rm d.o.f}\approx 0.9$) is perfectly consistent with the FSS
estimate $\beta_c=0.5091503(5)$.  The corresponding curve is drawn
in Fig.~\ref{ximcht}. An identical result, i.e.,  $\beta_c=0.5091504(4)$,
is found by fitting the MC data of $\chi$, which shows that the
agreement with the FSS analysis is not just by chance.  We performed
similar analyses also for $\lambda=2.2$, and in the case of the ddXY
model for $D=1.02$ and $D=1.20$. The results denoted by MC+HT are reported in
Table~\ref{betac}.  The comparison with the FSS estimates is overall
satisfactory.  This successful analysis should be contrasted with the
case of the standard XY model, where the fit of the MC results
in the HT phase by bIA1 does not provide acceptable results:
indeed, most of the approximants are defective. This
fact may be explained by the presence of sizable leading scaling
corrections.

In order to determine the critical exponents $\gamma$ and $\nu$, we
analyze the 22nd-order series of $\chi$ and 21st-order series of
$\xi^2/\beta$ respectively, using bIA$n$'s with $n=1,2,3$ biased at
the best available estimate of $\beta_c$.  Estimates of the exponent
$\eta$ can be obtained by using the scaling relation
$\eta=2-\gamma/\nu$. More precise estimates of the product $\eta\nu$
can be obtained by using the so-called critical-point renormalization
method (CPRM) applied to the series of $\chi$ and $\xi^2/\beta$, see
Ref.~\onlinecite{CHPRV-01} for details.  For the $\phi^4$ and ddXY models
we performed analyses at different values of the $\lambda$ and $D$ to
determine the dependence of the effective exponents on $\lambda$ and $D$
due to the residual leading corrections to scaling that are not taken into
account in the analysis.  In Table~\ref{biasedan} we report some
intermediate results, i.e., the results for each bIA$n$ analysis,
which will then lead us to our final estimates reported below. We
closely follow Ref.~\onlinecite{CHPRV-01}, so we refer to it for details. We
only mention that the set of bIA$n$'s that we consider are those with
$q=2$, $s=1/2$, $n_\sigma=2$ in the definitions reported in
Ref.~\onlinecite{CHPRV-01}.

\begin{table}
\squeezetable
\caption{Results for the exponents $\gamma$, $\nu$, and $\eta$
as obtained by using bIA$n$.
}
\label{biasedan}
\begin{ruledtabular}
\begin{tabular}{llllll}
\multicolumn{1}{c}{model}&
\multicolumn{1}{c}{$\beta_c$}&
\multicolumn{1}{c}{approximants}&
\multicolumn{1}{c}{$\gamma$}&
\multicolumn{1}{c}{$\nu$}&
\multicolumn{1}{c}{$\eta\nu$ from CPRM}\\
\colrule \hline
$\phi^4,\,\lambda=1.90$ & 0.5105799(7) &
bIA1 & 1.31718(3)[10] & 0.67116(10)[3] & 0.02531(4) \\

& & bIA2  & 1.31726(8)[12] & 0.67119(5)[3] & 0.02527(10) \\

& & bIA3  & 1.31723(3)[11] & 0.67118(14)[4] & 0.02528(5)  \\

$\phi^4,\,\lambda=2.07$ & 0.5093835(5) &
bIA1 & 1.31757(2)[7] & 0.67153(4)[3] & 0.02553(5) \\

& & bIA2  & 1.31759(4)[7] & 0.67155(6)[3] & 0.02552(12) \\

& & bIA3  & 1.31758(2)[7] & 0.67153(5)[3] & 0.02552(4)  \\

$\phi^4,\,\lambda=2.10$ & 0.5091504(4) &
bIA1 & 1.31767(3)[6] & 0.67160(3)[2] & 0.02557(5) \\

& & bIA2 & 1.31769(5)[5] & 0.67163(6)[2] & 0.02556(12) \\

& & bIA3 & 1.31767(3)[5] & 0.67160(4)[2] & 0.02556(5)  \\

$\phi^4,\,\lambda=2.20$ & 0.5083355(7) &
bIA1 & 1.31787(2)[8] & 0.67178(2)[5] & 0.02569(6) \\

& & bIA2 & 1.31789(5)[10] & 0.67181(7)[5] & 0.02567(12) \\

& & bIA3 & 1.31788(5)[9] & 0.67179(4)[4] & 0.02569(5)  \\

ddXY,$\,D=1.02$ & 0.5637963(4) &
bIA1     & 1.31757(15)[5] & 0.67141(4)[3] & 0.02547(5) \\

& & bIA2 & 1.31746(9)[4]  & 0.67143(6)[2] & 0.02550(23) \\

& & bIA3 & 1.31745(12)[5] & 0.67149(9)[3] & 0.02533(19)  \\

ddXY,$\,D=1.03$ & 0.5627975(13) &
bIA1 & 1.31767(15)[15] & 0.67148(4)[7] & 0.02550(4) \\

& & bIA2 & 1.31756(9)[14] & 0.67150(5)[6] & 0.02552(23) \\

& & bIA3 & 1.31755(10)[15] & 0.67155(8)[9] & 0.02537(18)\\

ddXY,$\,D=1.20$ & 0.5470388(11)
& bIA1 & 1.31871(5)[14] & 0.67227(8)[7]  & 0.02602(4) \\

& & bIA2 & 1.31872(11)[13] & 0.67232(10)[6] & 0.02597(28) \\

& & bIA3 & 1.31868(13)[13] & 0.67224(10)[8] & 0.02575(52)
\end{tabular}
\end{ruledtabular}
\end{table}

In the case of the $\phi^4$ model, using the results of the bIA$n$
analysis at $\lambda=2.10$ and $\lambda=2.20$, and assuming a linear
dependence on $\lambda$ in between, we obtain the IHT results
\begin{eqnarray}
&&\nu = 0.67161(4)[2] + 0.0018(\lambda-2.10),\label{expphi4l} \\
&&\gamma = 1.31768(3)[5] + 0.0021(\lambda-2.10),\nonumber\\
&&\eta\nu = 0.02556(5) + 0.0013(\lambda-2.10).\nonumber
\end{eqnarray}
The central value at $\lambda=2.1$ is taken from the bIA2 and bIA3
analyses, the number in parentheses is basically the spread of
the approximants at $\lambda=2.10$ using the central value of
$\beta_c$ (we use $\beta_c=0.5091504(4)$), while the number in
brackets gives the systematic error due to the uncertainty on
$\beta_c$. The dependence of the results on the chosen value of
$\lambda$ is estimated by assuming a linear dependence, and evaluating
the coefficient from the results for $\lambda=2.2$ and $\lambda=2.1$,
i.e., from the ratio $\left[ Q(\lambda=2.2) -
Q(\lambda=2.1)\right]/0.1$, where $Q$ represents the quantity at hand.
Consistent results are obtained by using the pair of values
$2.07,2.20$ or $1.90,2.10$ instead of $2.10,2.20$.  In the case of the
ddXY model, using the results of the bIA$n$ analysis at $D=1.02$ and
$D=1.20$, we obtain
\begin{eqnarray}
&&\nu = 0.67145(6)[2] + 0.0046 (D-1.02),\label{expddxyl}\\
&&\gamma = 1.31746(11)[5] + 0.0070(D-1.02),\nonumber\\
&&\eta\nu = 0.02547(14) + 0.0031(D-1.02).\nonumber
\end{eqnarray}

An alternative and more straightforward analysis of HT series is
represented by the matching method, which was applied in
Refs.~\onlinecite{MJHMJG-00,CPRV-02} for the two- and three-dimensional
Ising models.  The idea is to generate sequences of estimates by
fitting the expansion coefficients with their asymptotic form.  By
adding a sufficiently large number of terms one can make the
convergence as fast as possible, although, of course, the procedure
becomes unstable if the number of terms included is too large compared
to the number of available terms.  The estimates of the critical
exponents from this analysis are consistent with the bIA$n$ results
and with the fact that the $\phi^4$ model at $\lambda\approx 2.1$ and
the ddXY model at $D\approx 1.1$ are approximately improved, i.e., the
coefficients of the leading scaling corrections are consistent with
zero.  However, the results of this analysis are not sufficiently
stable and precise to improve the estimates already
obtained.

Estimates of the critical exponents can be then obtained by
evaluating Eqs.~(\ref{expphi4l}) and (\ref{expddxyl}) at the FSS
estimates of $\lambda^*$, i.e., $\lambda^*=2.15(5)$, and $D^*$,
i.e., $D^*=1.06(2)$, where the residual effect of the leading scaling
correction should vanish.  We refer to these results as MC+IHT
estimates.  For the $\phi^4$ theory we obtain
\begin{eqnarray}
&&\nu = 0.67170(4)[2]\{9\},\label{estphi4l}\\
&&\gamma = 1.31779(3)[5]\{11\}, \nonumber \\
&&\eta = 0.03816(7)\{10\} \nonumber
\end{eqnarray}
and using scaling and hyperscaling relations
\begin{eqnarray}
&&\alpha = 2-3\nu = -0.01510(12)[6]\{27\},
\label{hyestphi4l}\\
&&\eta = 2-\gamma/\nu = 0.03813(15)\{10\}.
\nonumber
\end{eqnarray}
The error due to the uncertainty on $\lambda^*$
is reported in braces.
For the ddXY model we obtain
\begin{eqnarray}
&&\nu = 0.67163(6)[2]\{9\},\label{estddxyl} \\
&&\gamma = 1.31774(11)[5]\{14\}, \nonumber\\
&&\eta = 0.03811(20)\{9\}, \nonumber
\end{eqnarray}
and
\begin{eqnarray}
&&\alpha = -0.01489(18)[6]\{27\},\label{hyestddxyl} \\
&&\eta = 2-\gamma/\nu = 0.03800(25)\{6\}.
\nonumber
\end{eqnarray}
There is good agreement between the MC+IHT estimates obtained from the
$\phi^4$ and ddXY models. We stress that this represents a nontrivial
check of the hypotheses underlying the IHT analysis. In particular,
this supports our working hypothesis that effects due to
next-to-leading nonanalytic corrections are negligible.  Finally,
estimates of the critical exponents $\delta$ and $\beta$ can be
obtained using the hyperscaling relations $\delta=(5-\eta)/(1+\eta)$
and $\beta=\nu (1+\eta)/2$.

\subsection{Combining IHT and FSS analyses}
\label{combFSSIHT}

\begin{figure}[tp]
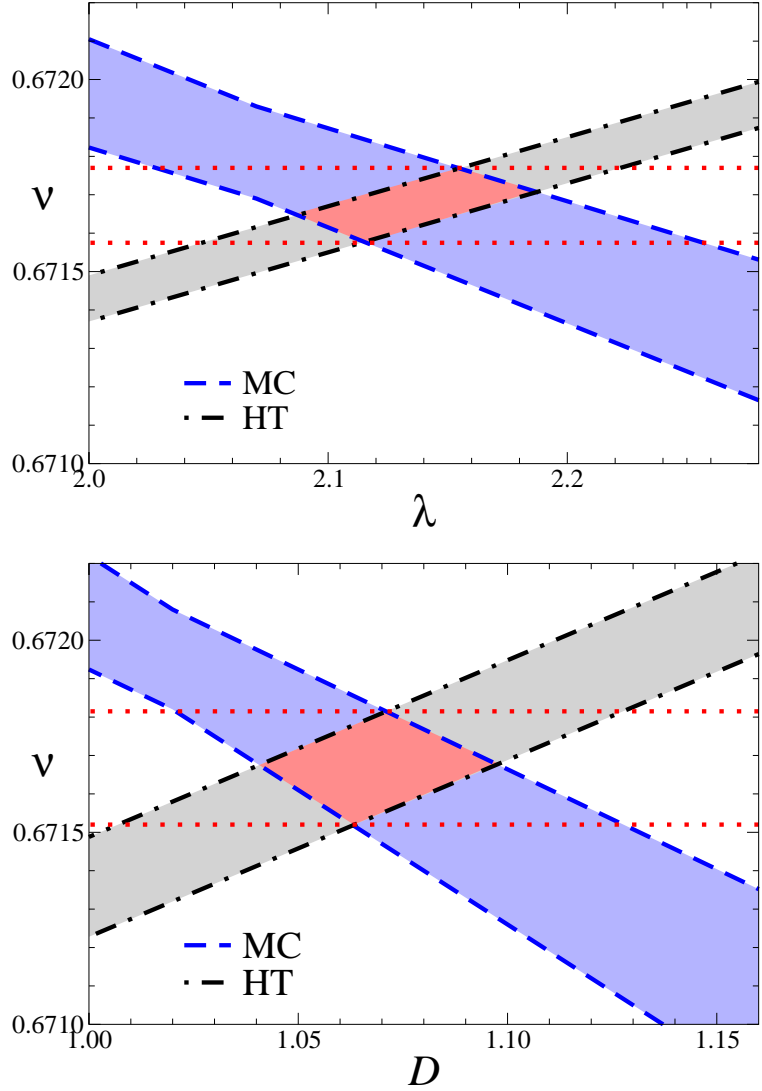

\centerline{\psfig{width=10truecm,angle=0,file=mcht.eps}}
\vskip 4mm
\centerline{\psfig{width=10truecm,angle=0,file=mchtd.eps}}
\vskip 2mm
\caption{\label{mcht}
Comparison of the HT and FSS results for the critical exponent
$\nu$ as functions of the 
parameters $\lambda$, $D$ around their optimal values.
The dotted lines correspond to the estimates of $\nu$ that are 
consistent with both the IHT and FSS analyses.
}
\end{figure}

Critical exponents can be also estimated by comparing
Eqs.~(\ref{expphi4l}) and (\ref{expddxyl}) with the analog FSS results
(\ref{fssla}) and (\ref{fssdi}) obtained from the MC data of $S_{U_4}$.
We note that the coefficients that give the dependence on $\lambda$ and $D$
in the IHT and FSS expressions
have opposite sign. Therefore, the results of the two
analyses agree only in a relatively small $\lambda$ (or $D$) interval. 
This is shown in
Fig.~\ref{mcht}.  This comparison provides an estimate of $\nu$
somewhat independent of the determination of the optimal values
$\lambda^*$, $D^*$.  Taking as
estimates of $\nu$ values consistent with both the IHT and FSS analyses, we
find the following FSS+IHT results
\begin{eqnarray}
&&\nu = 0.67168(10)\qquad \hbox{for the $\phi^4$ model}
,\label{phi4estcr}\\
&&\nu = 0.67167(15)\qquad \hbox{for the ddXY model} .
\label{ddxyestcr}
\end{eqnarray}
The good agreement between the two results nicely support our estimates of the
errors.  The FSS+IHT results represent our most precise estimates of the
critical exponent $\nu$. Corresponding estimates of $\alpha$ can be
obtained using the scaling relation $\alpha=2-3\nu$.

Note that this analysis also provides alternative estimates of $\lambda^*$ and
$D^*$. The intersection region in Fig.~\ref{mcht} indicates $\lambda^*=2.14(5)$
and $D^*=1.07(3)$, which are in good agreement with those obtained in
Sec.~\ref{ladstar}.

\subsection{Universal amplitude ratios}
\label{ratioofamp}

Using HT methods, it is possible to compute the first coefficients
$g_{2j}$ and $r_{2j}$ appearing in the small-magnetization
expansion of the Helmholtz free
energy and of the equation of state.\cite{PV-rev}
Indeed, these quantities can be expressed in terms of
zero-momentum $2j$-correlation functions and of the correlation
length. They are defined as
\begin{equation}
g_4 \equiv  - {3\over 2} {\chi_4\over \chi^2 \xi^3},
\label{grdef}
\end{equation}
and
\begin{eqnarray}
r_6 \equiv && 10 - {10\over 9}{\chi_6\chi_2\over \chi_4^2},
    \label{r2jgreen}\\
r_8 \equiv && 280 - {560 \over 9}{\chi_6\chi_2\over \chi_4^2}
+{35\over 27}{\chi_8\chi_2^2\over \chi_4^3}, \nonumber
\end{eqnarray}
etc...

Using their extended HT series,
we update the estimates of $g_4$, $r_6$, and $r_8$
obtained in Ref.~\onlinecite{CHPRV-01}.
Consider a universal amplitude ratio $A$ which, for
$t\equiv1 - \beta/\beta_c\to0$, behaves as
\begin{equation}
A(t) = A^* + c_1 t^\Delta + c_2 t^{\Delta_2} + \ldots + a_1 t + a_2 t^2+ \ldots
\end{equation}
In order to determine $A^*$ from the HT series of $A(t)$, we consider
bIA1's, whose behavior at $\beta_c$ is given by
\begin{equation}
{\rm IA1} \approx
f(\beta) \left(1 - \beta/\beta_c\right)^{\zeta} + g(\beta),
\label{IA1bh}
\end{equation}
where $f(\beta)$ and $g(\beta)$ are regular at $\beta_c$, except when
$\zeta$ is a nonnegative integer.  In particular,
\begin{equation}
\zeta = {P_0(\beta_c)\over P_1'(\beta_c)},\qquad\qquad
    g(\beta_c) = - {R(\beta_c)\over P_0(\beta_c)}.
\label{IA1bhf}
\end{equation}
In the case we are considering, $\zeta$ is positive and therefore,
$g(\beta_c)$ provides an estimate of $A^*$.  Moreover, for improved
Hamiltonians we expect $\zeta =\Delta_2 \approx 1.2$, instead of
$\zeta=\Delta \approx 0.53$.

In the case of $g_4$ we analyze the series $\beta^{3/2} g_4
=\sum_{i=0}^{20} a_i\beta^i$.  In Table~\ref{g4} we report some
results for several values of $\lambda$ and $D$.
Assuming a linear dependence on $\lambda$, $D$ around
their optimal values, we find
\begin{equation}
g_4 = 21.153(4) + 0.14 (\lambda-2.10) \label{g4res}
\end{equation}
for the $\phi^4$ model, and
\begin{equation}
g_4 = 21.15(3) + 0.4 (D-1.03)
\label{g4resd}
\end{equation}
for the ddXY model.
We estimate the critical value of $g_4$ by evaluating the
above expressions at $\lambda^*$ and $D^*$. We obtain respectively
\begin{eqnarray}
g_4 = 21.160(4)\{7\},\qquad
g_4 = 21.16(3)\{1\},
\end{eqnarray}
where the error in braces is related to the uncertainty on the estimates of
$\lambda^*$ and $D^*$.  We consider $g_4=21.16(1)$ as our final estimate.
This significantly improves our earlier result\cite{CHPRV-01}
$g_4=21.14(6)$ and is fully consistent with 
the FT estimate $g_4=21.16(5)$ obtained from an
analysis of six-loop perturbative series.\cite{GZ-98,gbar} Other results for
$g_4$ can be found in Ref.~\onlinecite{PV-rev}.  The results for the
nonanalytic exponent $\zeta$, reported in Table~\ref{g4}, give
\begin{equation}
\zeta= 1.16(6) + 1.3 (\lambda-2.10)
\label{zetares}
\end{equation}
in the case of the $\phi^4$ model.
Evaluating $\zeta$ at $\lambda=\lambda^*$ we obtain an estimate of
$\Delta_2$, i.e., $\Delta_2=1.23(6)\{7\}$, corresponding to
$\omega_2=1.83(19)$.  A consistent, but less precise, estimate can be obtained
from the results for the ddXY model.

It is worth mentioning some results obtained for the analysis of the
HT series of $g_4$ in the standard XY model.  Using bIA1's, biased so
that $\beta_c=0.4541652(11)$ and $\Delta=0.527(13)$ (corresponding to
$\omega=0.785(20)$), we obtain $g_4=21.12(5)$, which is in good
agreement with the estimate obtained from the improved models, but
much less precise.  Moreover, if we analyze the same series biasing
$\beta_c=0.4541652(11)$ and $g_4=21.16(1)$ and taking 
scaling-correction exponent $\Delta$ as a free parameter, 
we obtain the estimate
$\Delta=0.56(5)$, which is in agreement with the result obtained
from the FSS analysis.

\begin{table}
\squeezetable
\caption{
Estimates of the fixed-point value of $g_4$ from the 20th-order HT series
of $\beta^{3/2} g_4$.  The error due to the
uncertainty of $\beta_c$ is negligible.
}
\label{g4}
\begin{ruledtabular}
\begin{tabular}{lll}
\multicolumn{1}{c}{model}&
\multicolumn{1}{c}{$g_4$}&
\multicolumn{1}{c}{$\zeta$}\\
\colrule \hline
$\phi^4,\,\lambda=2.07$ & 21.149(5) & 1.13(6) \\
$\phi^4,\,\lambda=2.10$ & 21.153(4) & 1.16(6) \\
$\phi^4,\,\lambda=2.20$ & 21.167(4) & 1.28(7) \\
ddXY,$\,D=0.90$ & 21.13(7) & 0.9(2) \\
ddXY,$\,D=1.02$ & 21.15(3) & 1.1(4) \\
ddXY,$\,D=1.03$ & 21.15(3) & 1.1(3) \\
ddXY,$\,D=1.20$ & 21.22(2)  & 2.5(1.1) \\
\end{tabular}
\end{ruledtabular}
\end{table}

Similar analyses applied to the 18th-order HT series of $r_6$ and $r_8$
provide the estimates $r_6=1.96(2)$ and $r_8=1.5(1)$, which substantially
confirm those obtained in Ref.~\onlinecite{CHPRV-01}.  These results
can be used to compute approximations of the critical equation of
state, using the method
outlined in Refs.~\onlinecite{CPRV-00-2,CHPRV-01}, which is based on an
appropriate analytic continuation in the $t,H$ space.  Using our new estimates
of $\alpha$, $\eta$, $r_6$ and $r_8$, we obtain results for the critical
equation of state and universal amplitude ratios that are substantially
equivalent to those obtained in Ref.~\onlinecite{CHPRV-01},
essentially because our new HT results do not significantly improve
the estimates of $r_6, r_8$, and
no precise and reliable estimates of the higher-order coefficients $r_{2j}$
are available.  Therefore we do not provide further details.  We only report
the result $R_\alpha\equiv (1 - A^+/A^-)/\alpha=4.3(2)$ which is relevant for
the superfluid transition in $^4$He. For comparison, we mention the
FT result~\cite{SD-03} $R_\alpha=4.43(8)$, the numerical MC
results $R_\alpha=4.20(5)$ (Ref.~\onlinecite{CEHMS-02}) and
$R_\alpha=4.0(1)$ (Ref.~\onlinecite{martin-inprep}) 
and the experimental estimate
$R_\alpha=4.154(22)$ reported in Ref.~\onlinecite{Lipa-etal-03}.


\begin{thebibliography}{99}
\bibitem{G05CAFgen} 
The G05CAF generator is a linear congruential generator:
$I_{j+1} = {\rm Mod} [a I_j + c,n]$ with $a=13^{13}$, $c=0$ and
$n=2^{59}$.
\bibitem{gbar}
In FT calculations, such as those reported in  Ref.~\onlinecite{GZ-98}, one
usually introduces the rescaled coupling $\bar{g}\equiv 5 g_4/(24\pi)$.  Our
new estimate of $g_4$ corresponds to $\bar{g}=1.4032(7)$, to be compared
with the FT result $\bar{g}=1.403(3)$ (Ref.~\onlinecite{GZ-98}).
\end{thebibliography}
\end{document}